\documentclass[12pt,preprint]{aastex}

\newcommand{\myemail}{pteixeira@cfa.harvard.edu}

\shorttitle{From Filament to Cores to Stars in Lupus\,3}
\shortauthors{Teixeira et al.}

\received{2005 January 28}
\begin{document}

\title{From Dusty Filaments to Cores to Stars:\\ An Infrared Extinction Study of \object{Lupus\,3}\altaffilmark{1}}

\author{Paula S. Teixeira \altaffilmark{2,3} and Charles J. Lada}
\affil{Harvard-Smithsonian Center for Astrophysics, 60 Garden Street, Cambridge, MA, 02138, USA}
\email{\myemail, clada@cfa.harvard.edu}
 \and 
\author{Jo\~ao~F. Alves}
\affil{European Southern Observatory, Karl-Schwartzschild-Strasse 2, D-85748 Garching bei M\"unchen, Germany}
\email{jalves@eso.org}

\altaffiltext{1}{Based on observations carried out at ESO, La Silla and Paranal, Chile.}
\altaffiltext{2}{Departamento de F\'{\i}sica da Faculdade de Ci\^encias da Universidade de Lisboa, Portugal}
\altaffiltext{3}{Centro de F\'{\i}sica Nuclear da Universidade de Lisboa, Portugal}

\begin{abstract}
We present deep near-infrared (NIR) observations of a dense region of the Lupus\,3 cloud obtained with the European Southern Observatory's (ESO) New Technology Telescope (NTT) and Very Large Telescope (VLT). Using the near-infrared color excess (NICE) method we cons\-truct a detailed high angular re\-so\-lu\-tion dust extinction map of the cloud. The dust extinction map reveals embedded globules, a dense filament, and a dense ring structure. We derive dust column densities and masses for the entire cloud and for the individual structures therein. We construct radial extinction profiles for the embedded globules and find a range of profile shapes from relatively shallow profiles for cores with low peak extinctions, to relatively steep profiles for cores with high extinction. Overall the profiles are similar to those of pressure truncated isothermal spheres of varying center-to-edge density contrast. We apply Bonnor-Ebert analysis to compare the density profiles of the embedded cores in a quantitative manner and derive physical parameters such as temperatures, central densities, and external pressures. We examine the stability of the cores and find that two cores are likely stable and two are likely unstable. One of these latter cores is known to harbor an active protostar. Finally, we discuss the relation between an emerging cluster in the Lupus\,3 cloud and the ring structure identified in our extinction map. Assuming that the ring is the remnant of the core within which the cluster originally formed we estimate that a star formation efficiency of approximately 30\% characterized the formation of the small cluster. Our observations of the Lupus\,3 cloud suggest an intimate link between the structure of a dense core and its state of star forming activity. The dense cores in this cloud are found to span the entire range of evolution from a stable, starless core of modest central concentration, to an unstable, star-forming core which is highly centrally concentrated, to a significantly disrupted core from which a cluster of young stars is emerging.
\end{abstract}

\section{Introduction}
\label{sec:introd}

Knowledge of the initial conditions is essential to developing a theoretical description of the process of star formation. This is the key motivation for probing the structure of clouds, since their density structure and the evolution of this structure to produce stars is pivotal to the physical process of star formation.  Earlier studies have shown that much can be learned about the structure of clouds from sensitive infrared extinction measurements of stars background to the clouds.  Infrared extinction studies of filamentary clouds such as \object{IC\,5146} \citep{ic5146}, \object{L\,977} \citep{l977a,l977b} and \object{L\,781} \citep{l781} have indicated that such elongated clouds are characterized by well behaved radial density profiles. Very deep near-infrared $J$, $H$, and $K_s$ observations of isolated cores and globules such as \object{B\,68} \citep{alves01}, \object{B\,335} \citep{b335} and \object{Coalsack Globule 2} \citep{lada04} have produced detailed maps of their structure and radial density profiles with relatively high spatial resolution (10\arcsec\ to 15\arcsec).  Moreover, these measurements produced exquisitely sampled radial density profiles of the globules that display qualitative and quantitative differences in shape (steepness) and these differences appear to be correlated with the evolutionary state of the clouds \citep[see also][]{racca}. These studies have provided a detailed picture of the structure and physical conditions both in isolated protostellar and starless (and
presumably pre-protostellar) globules.

Most stars, however, form in more massive cloud complexes, in embedded clusters within GMCs and not in isolation.  Thus, it would be very useful to obtain knowledge of the structure of such clouds in similar detail.  Due to extremely high extinctions and rich embedded populations, it is generally not possible to obtain detailed extinction maps of massive, cluster-forming cores even with the deepest near-infrared surveys.  However, it is possible to obtain detailed observations of certain smaller and less massive filamentary cloud complexes such as IC\,5146 and L\,977 which are intermediate in their properties between isolated globules and GMCs.  These clouds typically exhibit significant structure, containing dense cores of molecular gas that span a range in mass.  These cores are capable of producing both single stars (similar to isolated globules) and small stellar groups and poor clusters (similar to GMCs).  Unfortunately previous extinction observations of such clouds were not sensitive enough to resolve in detail the structure of the individual cores within the clouds.  Deeper near-infrared surveys of such clouds can significantly improve on this situation. In an attempt to meet the goal of producing deep near-infrared extinction maps of a filamentary cloud, we have obtained a deep near-infrared survey of the highest extinction regions of Lupus\,3 dark cloud complex, a nearby star-forming cloud complex in the southern hemisphere. 

The Lupus\,3 complex is an active region of star formation, containing 27 PMS objects previously identified in H$\alpha$ \citep{schwartz} and near-infrared studies \citep{comeron} in our surveyed area, one of which is a Herbig Ae star \citep{the78}. An HH object \citep{the62, krautter97, nakajima} is also identified in this region. Of the 27 PMS objects, roughly 17 are concentrated in a small cluster in the western portion of the cloud. There is no clear consensus on the distance to Lupus\,3.  The distances cited in the li\-te\-ra\-ture range from 100\,pc \citep{knude} to $223$\,pc, the averaged \emph{Hipparcos} distances of the stars HR\,5999 and HR\,6000 \citep{esa}.  We adopt the intermediate value of 140\,pc derived by \citet{hughes} for this study. The internal structure of Lupus\,3 was first discussed by \cite{gf21} who listed it as GF\,21 in their catalogue of clouds they classified as globular filaments. Subsequent studies have probed its large scale structure \citep[e.g.][]{andreazza}. More recently, \citet{nakajima03} have produced a deep near-infrared extinction map of the darkest region of the cloud. They resolved two very dark, moderately massive cores, both with a catalogued IRAS source, although only one shows independent evidence for star formation.  These observations suggest that it would be fruitful to produce a sensitive extinction map of a larger region of the Lupus\,3 complex.

In this paper we present results of a deep near-infrared imaging survey of the Lupus\,3 complex performed with the ESO NTT and VLT telescopes.  We have used the NICE method \citep{lada04} to derive precise extinctions to thousands of stars behind the complex.  This is the first time the NICE method is applied to derive extinctions for the whole Lupus\,3 cloud with high spatial resolution. From this data we are able to construct detailed spatial extinction maps of the cloud and we achieved sufficient sensitivity and angular resolution to enable construction of the radial density profiles for some of the individual cores within the filamentary cloud.

We describe the observations and the data reduction techniques in \S\, \ref{sec:obs}. The results are presented in \S \ref{sec:results}  followed by the analysis and the discussion on the evolution of the cores and star formation in Lupus\,3 in \S \ref{sec:analysisDiscuss}.

\section{Observations and Data Reduction}
\label{sec:obs}

\begin{figure}
\centering
\includegraphics[width=\textwidth]{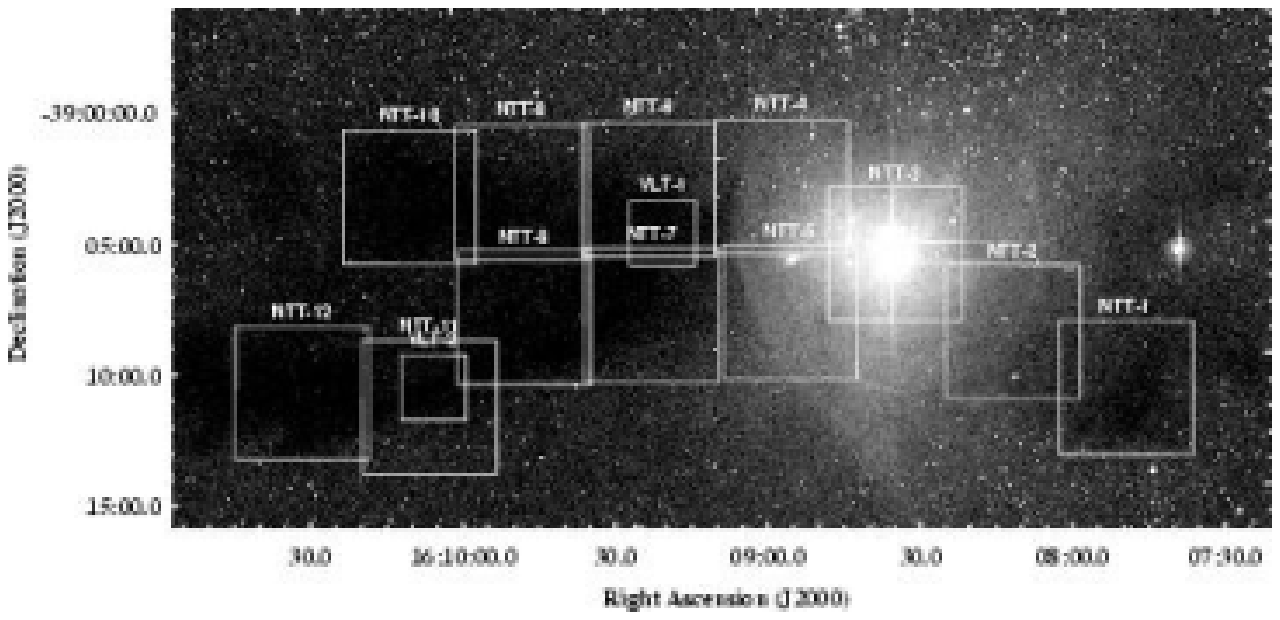}
\caption{Optical image of the Lupus\,3 molecular cloud obtained from the blue Digitized Sky Survey, POSS2. Our NTT surveyed area is marked by the bigger boxes while the two regions imaged with VLT are shown in smaller boxes.}
\label{fig:survey}
\end{figure}

Observations of the Lupus\,3 filamentary cloud were carried out in March 2000 using the SofI camera at European Southern Observatory's (ESO) New Technology Telescope (NTT) in La Silla, Chile. SofI is equipped with a 1024$\times$1024 pixel Hawaii HgCdTe array. The image scale of the array is 0.295\arcsec\, per pixel, giving a field of view of $\sim$\,5\arcmin$\times$5\arcmin. We obtained 12 frames both in the $H$--band and $K_s$--band filters and 3 of those frames were also observed with the $J$--band filter to better analyze star formation in those regions. Total integration times at $J$, $H$, and $K_s$ for each frame is 300\,s, 500\,s, and, 180\,s, respectively. The frame that contains the cluster was observed for 300\,s in all three filter bands. The survey area extends from 16$^\mathrm{h}$07$^\mathrm{m}$36$^\mathrm{s}$ to 16$^\mathrm{h}$10$^\mathrm{m}$43.2$^\mathrm{s}$, and from -39\degr15\arcmin00\arcsec\, to -38\degr57\arcmin36\arcsec, as shown in Figure \ref{fig:survey}. This area corresponds to the dark filaments of Lupus\,3, as well as the associated cluster, so that we would have areas with and without star formation in our survey. A control field was also observed 18\arcmin\, north of the Lupus\,3 cluster in $J$, $H$, and $K_s$ with total integration times are 90\,s, for each filter band. For two regions that are heavily extincted, and as such were not penetrated by the deep NTT observations, deeper observations were taken in July 2000 using ISAAC on ESO's Very Large Telescope (VLT) in Paranal, Chile in $H$ and $K_s$--bands. ISAAC is also equipped with a Hawaii 1024$\times$1024 pixel HgCdTe array, having a pixel scale of 0.148\arcsec\, per pixel and giving a field of view of 2.5\arcmin$\times$2.5\arcmin. Integration times at $H$ and $K_s$ for each frame is 3600\,s and 1800\,s, respectively. After dark subtracting and dividing by the flat-field, we corrected for crosstalk and linearity before combining the images for each field and band. The individual dithered images were reduced with an external IRAF\footnote{IRAF is distributed by the National Optical Astronomy Observatories, which are operated by the Association of Universities for Research in Astronomy, Inc., under cooperative agreement with the National Science Foundation.} package named DIMSUM (Deep Infrared Mosaicing Software) that performs a very good sky-subtraction in two passes, producing a final image that has very low background noise. 
Source extraction photometry was then performed on the final images using the FLAMINGOS PinkPhot pipeline\footnote{this work was supported in part by NSF grants, AST97-3367 and AST02-02976 to the University of Florida} which does automated PSF fitting photometry. Photometric calibration was done using 2MASS photometry to determine zero points. Astrometry was also calibrated with 2MASS and has an average uncertainty of 0.14\arcsec. The final FWHM of point sources in the final images is summarized in Table \ref{tab:obs} along with the number of stars per field in each band and completeness limits.

\section{Results}
\label{sec:results}

\subsection{The Dust Extinction Map}
\label{subsec:build_map}

We use the NICE method \citep{lada94} to derive extinctions from the individual measured color excesses of extincted stars in the target field. Briefly, the color excess is given by:
\begin{equation}
\centering
E(H-K_s)=(H-K_s)_\mathrm{observed}-(H-K_s)_\mathrm{intrinsic}
\label{eq:excess}
\end{equation}
where $(H-K_s)_\mathrm{intrinsic}$ is the intrinsic color of the background stars which is derived from averaging the $(H-K_s)$ colors of stars in a control field. The control field was observed using the NTT in the $J$, $H$, and $K_s$ filter bands. It is located 18\arcmin\, north of the Lupus\,3 cluster and at the same galactic la\-ti\-tude. The average $(H-K_s)$ colors of stars in the control field is calculated to be 0.19\,$\pm$\,0.11 magnitudes. \citet{nakajima03} derives $<(H-K_s)>=0.15$ magnitudes for a reference field 29\arcmin\, west of the Lupus\,3 cluster. The difference in the color excesses derived using these two values corresponds to 0.04 magnitudes, which is less than our statistical uncertainty in the color of our control field. 

We convert color excesses to equivalent visual extinction adopting a reddening law from \citet{av}:
\begin{equation}
\centering
A_{V}=15.93\,E(H-K_s)
\label{eq:av}
\end{equation}

\begin{figure}
\centering
\includegraphics[angle=90,width=\textwidth]{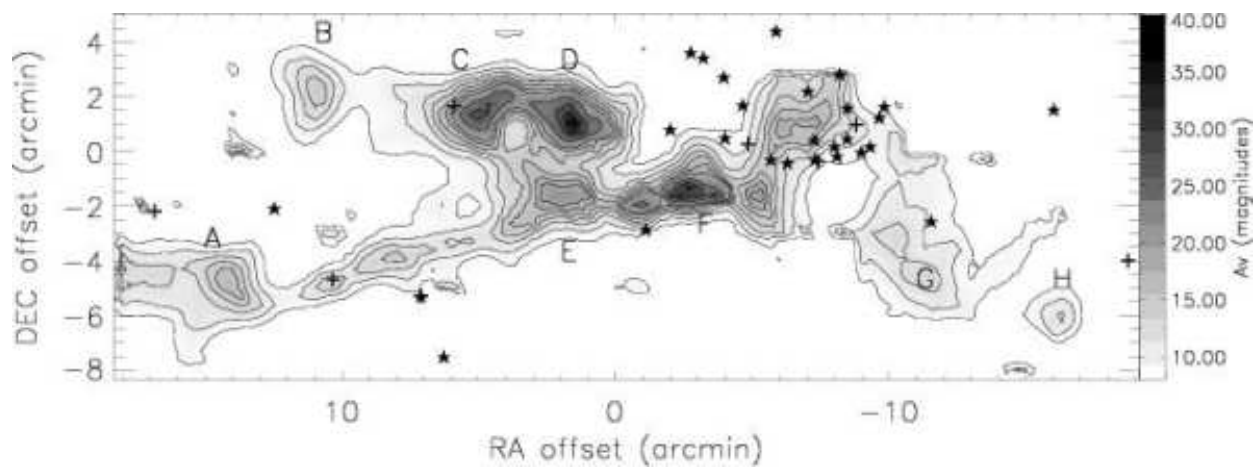}
\caption{Dust extinction map of the Lupus\,3 cloud constructed from deep NTT observations and using a Nyquist sampled gaussian spatial filter with 40\arcsec\, FWHM  (color version available in the electronic publication). Cores and filaments discussed in the text are labeled. The (0,0) position is (16$^\mathrm{h}$09$^\mathrm{m}$18.8$^\mathrm{s}$,-39\degr04\arcmin48.7\arcsec)(J2000). The extinction contours start at 8 magnitudes of visual extinction, increasing in steps of 2 until 20 and then in steps of 5 until 40 magnitudes.
The stars mark the positions of H$\alpha$ stars from \citet{schwartz} and the plus signs represent catalogued IRAS sources (see text for details).}
\label{fig:30map}
\end{figure}

The \citet{nyquist} sampled dust extinction map derived from our NTT and VLT observations, shown in Figure \ref{fig:30map}, was built by averaging the $(H-K_s)$ colors of the stars within a gaussian spatial filter with a FWHM of 30\arcsec\ and at 15\arcsec\ intervals. For a few dense regions where the deep NTT observations were not able to detect background stars through the cloud we have obtained additional VLT observations\footnote{We note that even though the VLT observations are deeper than the NTT observations, we only detect the intrinsically brighter sources of the VLT field because the stars are highly reddened. The background stars detected in the reddened VLT field are therefore similar to those in the unreddened NTT control field, making the latter appropriate to use for estimating intrinsic stellar colors of the stars in the VLT field.} Nonetheless with 30\arcsec\, resolution, there are still regions where the extinction is high enough to create empty pixels in the map, i.e, locations where no background stars are observed within a resolution element. In these cases we averaged the values of neighboring pixels to derive a lower limit to the extinction. To filter out foreground stars which would lower the true extinction in the map, we use a filter that excludes a star with an A$_\mathrm{v}$ less than 2 magnitudes within a pixel of average A$_\mathrm{v} \ge$ 10 magnitudes. In this way we have found an average foreground stellar density of 0.4 stars per pixel.
For regions of lower extinction we have many stars in a given pixel so foreground stars do not contribute significantly to the final averaged value. As an example, for a level of extinction of 6 magnitudes we find a mean number of 15 stars within a beam.

\begin{figure}
\centering
\includegraphics[angle=90]{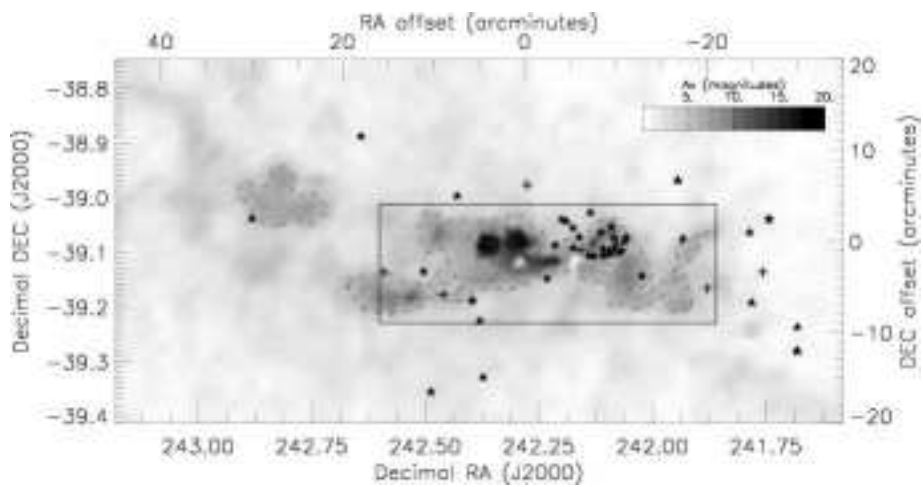}
\caption{Near-infrared dust extinction map built using a 2\arcmin\ square filter, Nyquist sampled, and 2MASS data of the Lupus\,3 region. The dashed lines corresponds to a level of A$_\mathrm{V}$ of 6 magnitudes, the stars mark the positions of H$\alpha$ stars from \citet{schwartz}, and the plus signs represent catalogued IRAS sources.}
\label{fig:2mass}
\end{figure}

As can be seen in Figure \ref{fig:30map}, the bordering regions of our NTT survey have extinctions greater than 8 magnitudes. To present the reader with a better idea of the lower extinction regions of the Lupus\,3 cloud, we applied the NICE method to build a 2MASS dust extinction map, shown in Figure \ref{fig:2mass}. To avoid a large number of holes in the map, we used a 2\arcmin\ square beam, spatially sampled every 1\arcmin. For reference we place the contour of 6 magnitudes of extinction (dashed line) which clearly outlines the denser part of Lupus\,3. Our NIR NTT and VLT data are located within the rectangle drawn in a solid line in this figure.

From our high resolution dust extinction map, shown in Figure \ref{fig:30map}, we can resolve several cores, and label  the most prominent with letters A to H. Our criteria for selecting a core consists of 
finding extinction peaks with A$_V >$\,8 magnitudes and at least 3 closed contours spaced by intervals of 2 magnitudes. The five most circular cores (A, B, C, D and H) are analyzed in this paper. 
There is a narrow filament that connects core A to core E and perhaps F. The filament seems to have in it two small elongated embedded cores that we shall not discuss here. The western end of this filament extends into core F, and beyond F we detect a structure that appears to be a ring of dust that borders a young cluster of PMS stars. Finally at the westernmost edge of the cloud there two lower density structures (G and H).
The cores we selected (A, B, C, D and H) are also seen in the 2MASS map of Figure \ref{fig:2mass}.

\subsection{Masses}
\label{subsubsec:mass}

We determine the gas co\-lumn density for each pixel in the dust extinction map, assuming here the standard gas-to-dust ratio \citep[N$_\mathrm{H}=1.9\,\times\,10^{21}\,$A$_V$ cm$^{-2}$, ][]{lilley55,jenkins74,bohlin} is also valid for the Lupus\,3 molecular cloud.
To calculate the total mass of the material within our surveyed region of Lupus\,3 we constructed a Nyquist sampled dust extinction map with square pixels, and integrated the extinction over the map:
\begin{equation}
M=2.2\times10^{-3}\left(\frac{beamsize}{30\arcsec}\right)^2 \left(\frac{d}{140}\right)^2 \sum_{i}^{n}\,A_V(i)\ (M_\sun)
\label{eq:mass}
\end{equation}

The mass calculated for the whole Lupus\,3 cloud from our NTT and VLT data is 84.7\,$\pm$\,0.8\,M$_\sun$. The error stated refers to the statistical uncertainty in the mass calculation, however, there are also systematic uncertainties present, for example the choice of map resolution in the mass calculation. To estimate this contribution, we calculated the mass for resolutions ranging from 30\arcsec\ to 70\arcsec. For the highest resolution maps, we have empty pixels and therefore underestimate the mass, for example, we derive 80.8\,M$_\sun$ for the 30\arcsec\ map. As we decrease the resolution, the empty pixels get filled in and the mass calculated increases. We find a maximum of 84.7\,M$_\sun$ for the 50\arcsec\ map. As we continue to decrease the resolution, the derived mass slowly begins to decrease because we are diluting the contribution from the highest density regions with neighboring lower density regions, e.g. for the 70\arcsec\ map we get 84.2\,M$_\sun$. This systematic uncertainty thus corresponds to 3.9\,M$_\sun$ or 5\% of the total mass calculated. Of course the largest source of systematic uncertainty is due to the uncertainty of the distance to the cloud ($\sim$\,50\%).
We have additionally calculated a mass of 194.2\,M$_\sun$ for the area of the 2MASS dust extinction map, shown in Figure \ref{fig:2mass}, that has A$_\mathrm{V}$ $>$ 2 magnitudes.

We calculate the masses of the individual cores embedded in the cloud in two ways from the NTT/VLT data. First, we use Equation \ref{eq:mass} to derive the total line of sight mass to each core. Second, we subtract a local plateau value between 4 - 9 magnitudes (see Table \ref{tab:masses} for the exact plateau levels for each core) from each pixel to account for the fact that the cores are embedded in a global fi\-la\-men\-ta\-ry structure and we wish to determine the mass of the cores by accounting for the contribution of the underlying filament. For this case, we use:
\begin{equation}
M=2.2\times10^{-3}\left(\frac{beamsize}{30\arcsec}\right)^2 \left(\frac{d}{140}\right)^2 \sum_{i}^{n}\,(A_V(i)-A_{V_\mathrm{base}})\ (M_\sun)
\label{eq:mass2}
\end{equation}
where A$_{V_\mathrm{base}}$ is the plateau and A$_V(i)\,\ge\,$A$_{V_\mathrm{base}}$.

The masses estimated for the whole Lupus\,3 cloud and the individual structures therein are given in Table \ref{tab:masses}, as well as the value for the plateau used to account for the contribution of the filamentary cloud to each core. In this table we also include the Jeans mass for each core, $M_\mathrm{J}=17\,T^{3/2}\,\bar{n}^{-1/2}\, M_\sun$, which we calculate from the average number density, $\bar{n}$, given by \hbox{$\bar{n}=3\,M/(4\,\mu\,m_\mathrm{H}\,\pi\,R^3)$}, where $M$ is the adjusted core mass (after subtraction of an appropriate plateau level of extinction), $\mu$ is the mean atomic weight per H atom, 1.36, and $m_\mathrm{H}$ is the H atomic mass. We also assume a temperature, $T$, of 10\,K. The radius, $R$, used in this calculation is obtained from a radial profile fitting as explained below in \hbox{Section \S\, \ref{subsubsec:profiles}}.

Masses for the cloud also have been estimated by previous observers from CO observations \citep[e.g.][]{murphy, hara, vilas-boas2}. \citet{hara}, from their C$^{18}$O observations, find a mass of 26\,M$_\sun$\ for their core 27 which would correspond roughly to our cores C, D, E, F and the ring. For the whole C$^{18}$O-emitting region of Lupus\,3 they calculate a mass of 105\,M$_\sun$\ which includes part of the cloud that lies to the northeast of our surveyed area (seen in our 2MASS extinction map in Figure \ref{fig:2mass}). The observations reported in this paper are consistent with the previous work mentioned, even considering that the area surveyed by CO observations is usually bigger, the resolution is lower and affects of uncertain chemistry (e.g. depletion) may lead to over- or underestimates of the mass.

\subsection{Radial Density Profiles of the Cores}
\label{subsubsec:profiles}

To study the selected cores (A, B, C, D, and H) we build individual dust extinction maps for each of them, with a higher resolution where possible. These zoomed in maps help us determine the location of the peak extinction of the core which we use as the center to build an averaged radial extinction profile. The profiles are built by constructing Nyquist sampled concentric circular 10\arcsec\, bins. We now briefly describe the dust extinction map and radial profile for each core.

\begin{figure}
\centering
\includegraphics[angle=90,scale=0.35]{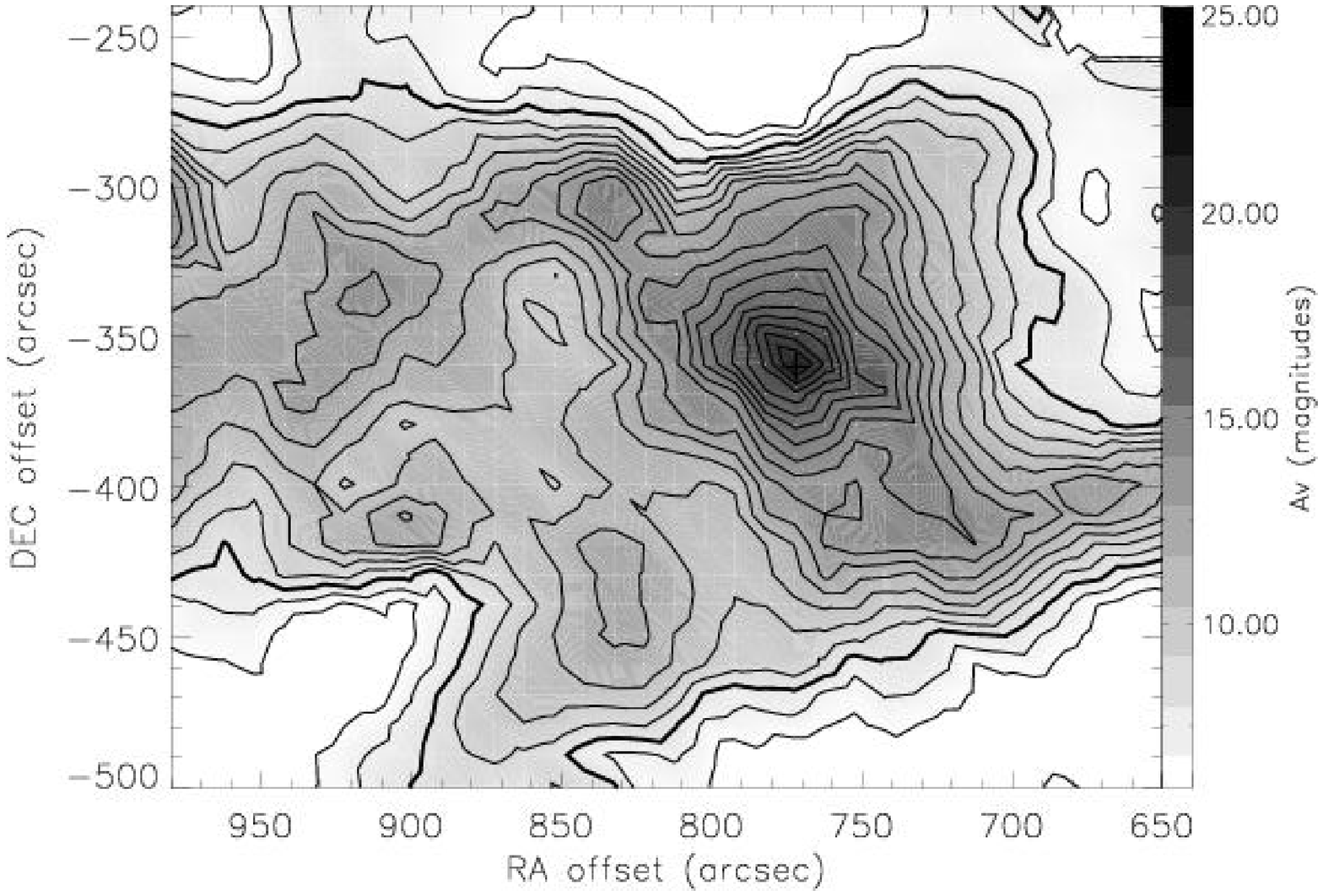}
\includegraphics[angle=90,scale=0.3]{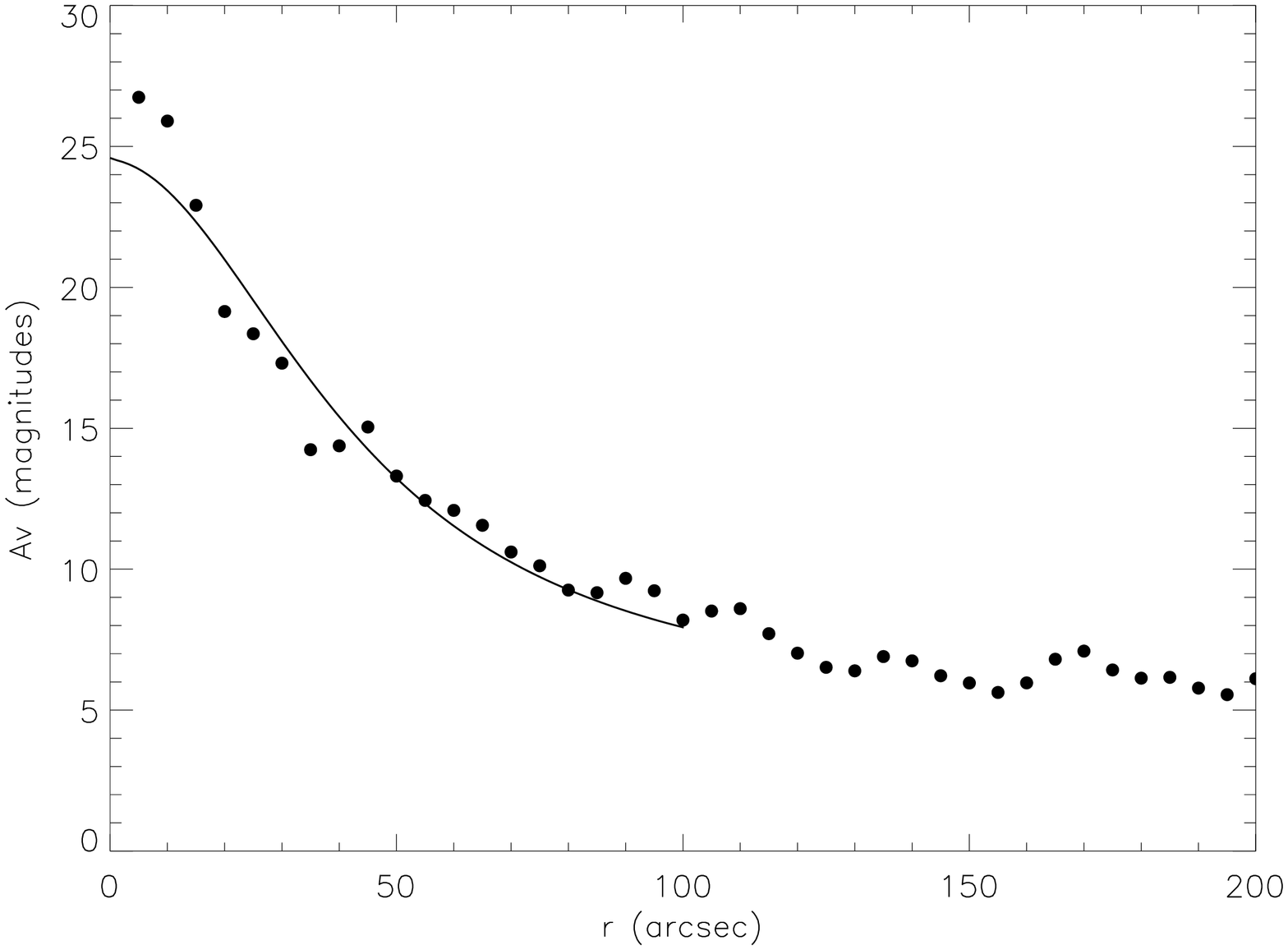}
\caption{\emph{left} 20\arcsec\, resolution dust extinction map of core A, embedded in the Lupus cloud. The contours range from 6 to 25 magnitudes in steps of one magnitude of visual extinction, the thicker contour corresponds to 8 magnitudes (color version available in the electronic publication); \emph{right} Extinction radial profile of core A, 10\arcsec\, resolution, Nyquist sampled. The overplotted line corresponds to a Bonnor-Ebert fit, discussed in \S\, \ref{subsec:properties}.}
\label{fig:coreAcont}
\end{figure}

The left panel of Figure \ref{fig:coreAcont} shows a 20\arcsec\, resolution dust extinction map for core A. The contours range from 6 to 25 magnitudes of visual extinction and the thicker contour corresponds to 8 magnitudes. There is no well defined boundary of the core, except for the denser region where the extinction is greater than 12 magnitudes. The right panel shows the corresponding extinction radial profile for core A. The position of the central condensation in the map is used as the center of the profile, which peaks at 27 magnitudes. The profile drops until it reaches a relatively constant level of 8 magnitudes. We define the radius of the core as the distance for which the profile reaches a constant level or \emph{plateau}. Core A has then a radius of 100\arcsec. The solid line overplotted corresponds to a Bonnor-Ebert fit that will be discussed in \S \ref{sec:analysisDiscuss}.

\begin{figure}
\centering
\includegraphics[angle=90,scale=0.35]{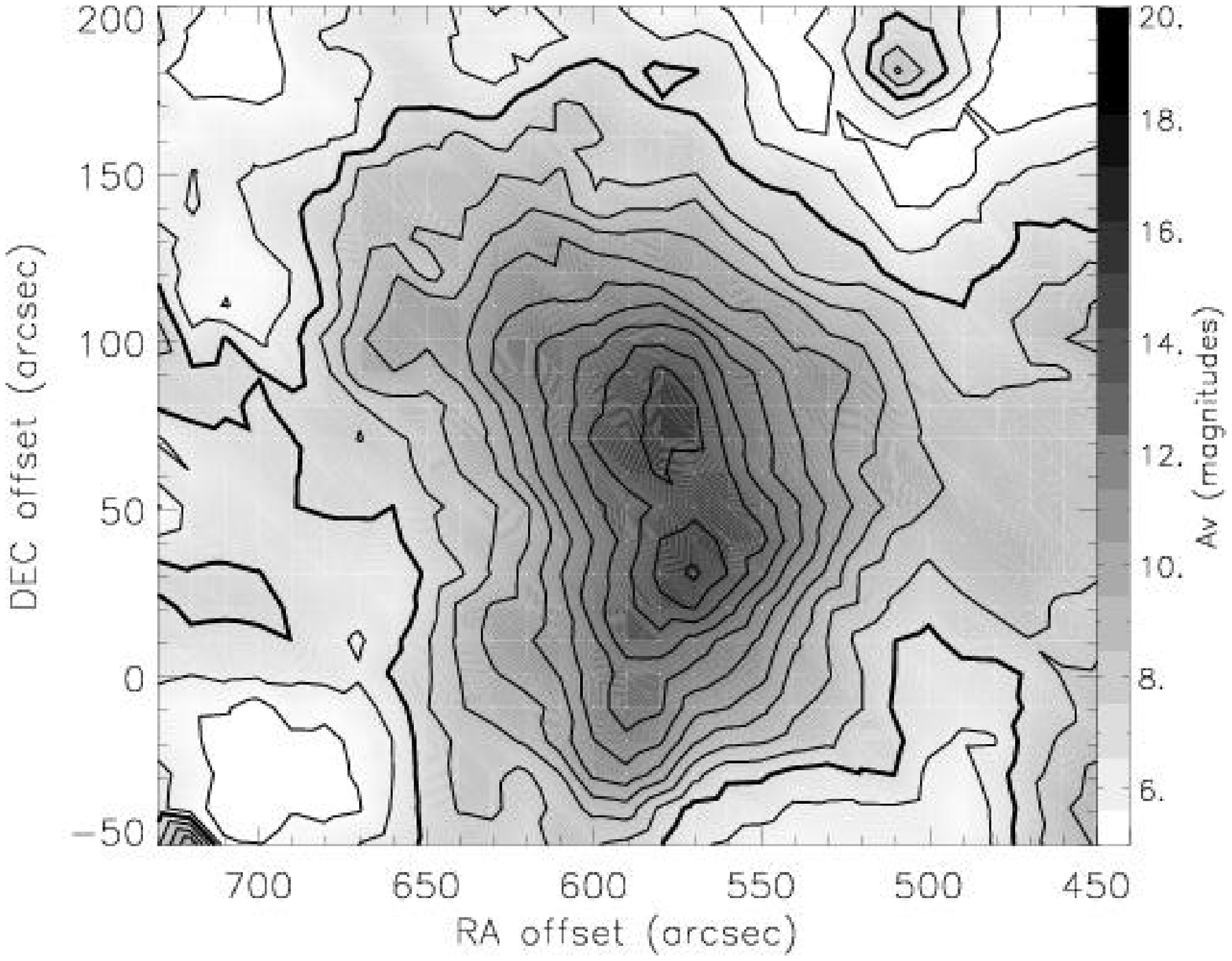}
\includegraphics[angle=90,scale=0.3]{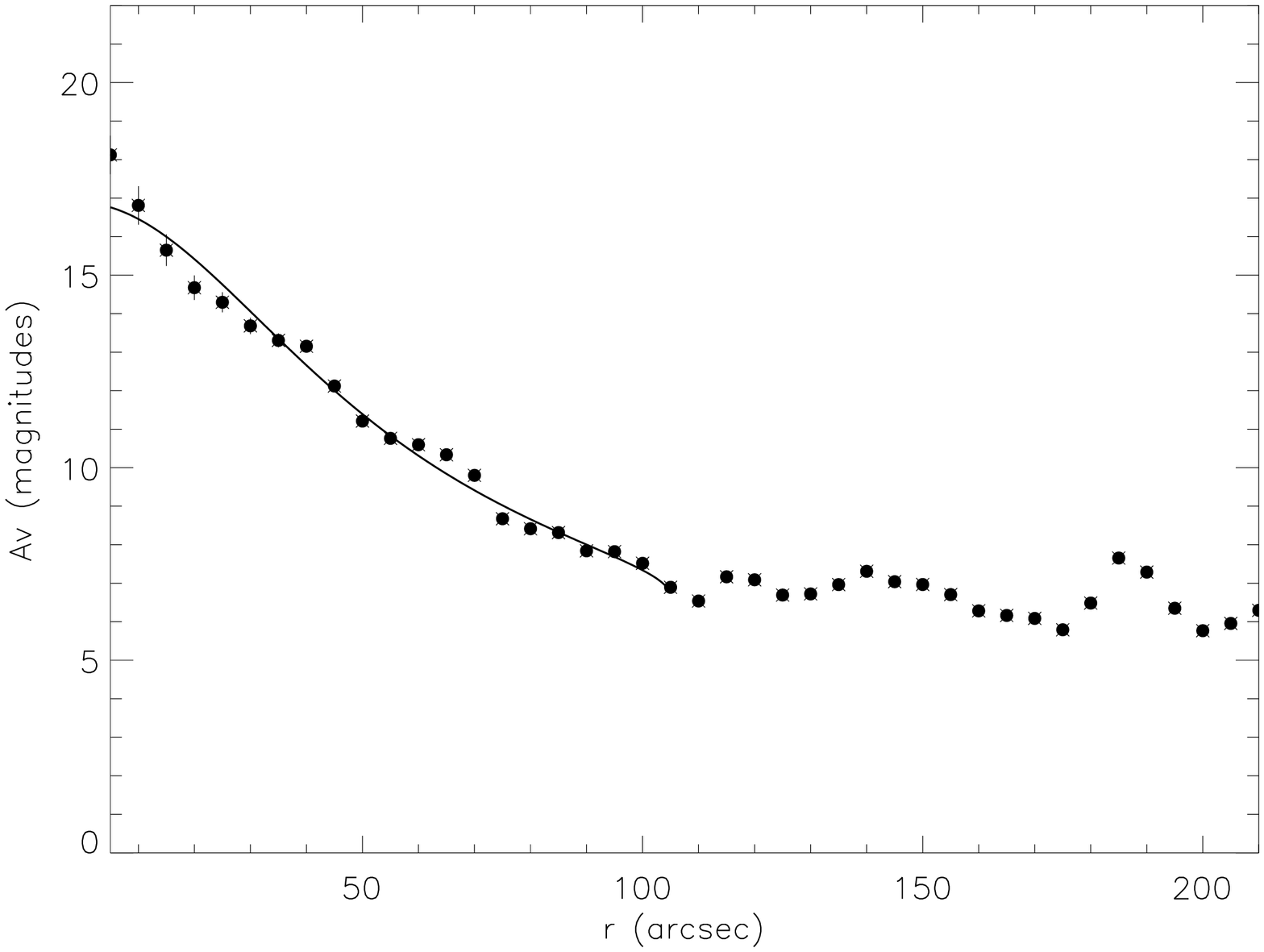}
\caption{\emph{left} 20\arcsec\, resolution dust extinction map of core B, embedded in the Lupus cloud. The contours range from 5 to 17 magnitudes in steps of one magnitude of visual extinction, the thicker contour corresponds to 7 magnitudes (color version available in the electronic publication); \emph{right} Extinction radial profile of core B, 10\arcsec\, resolution, Nyquist sampled. The overplotted line corresponds to a Bonnor-Ebert fit, discussed in \S\, \ref{subsec:properties}.}
\label{fig:coreBcont}
\end{figure}

Figure \ref{fig:coreBcont} shows the extinction map and radial profile for Core B. The 20\arcsec\, resolution dust extinction map (here contours range from 5 to 17 magnitudes and the thicker contour corresponds to 7 magnitudes) shows that core B is more isolated than core A with a small low extinction ``arm'' on the east. The central part of the core appears to break into two peaks. Care should be taken in interpreting this because the densest region has fewer stars in each pixel of the map (small number statistics) and the central division occurs where no background stars are present. This core has a better defined boundary than core A, however it is still not clear from the map what the exact size of the core is. As before, we use the extinction radial profile to determine the radius of the core. Choosing the radius of the core from the profile is a process that also has uncertainties (related to the adopted value of the background level, which is not unambiguous) though less than than those resulting from using the map. The profile peaks at 16.8 magnitudes and falls until it reaches a constant value of 7 magnitudes. We choose a radius of 100\arcsec\, for core B. 

\begin{figure}
\centering
\includegraphics[angle=90]{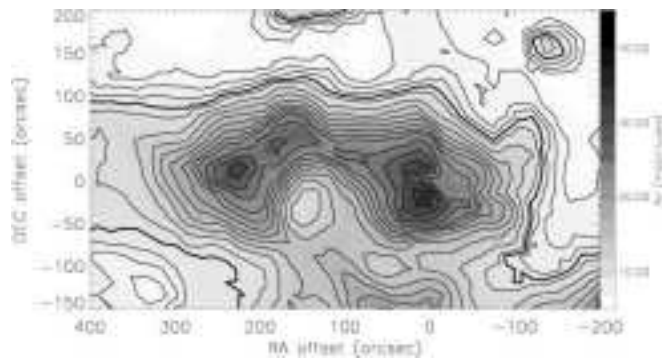}
\caption{30\arcsec\, resolution map of cores C and D. The contours range from 6 to 10 magnitudes in steps of 1 magnitude, to 40 in steps of 2 magnitudes and to 45 in a step of 5 magnitudes. The thicker contour corresponds to 9 magnitudes  (color version available in the electronic publication).}
\label{fig:coreCDcont}
\end{figure}

\begin{figure}
\centering
\includegraphics[angle=90,scale=0.31]{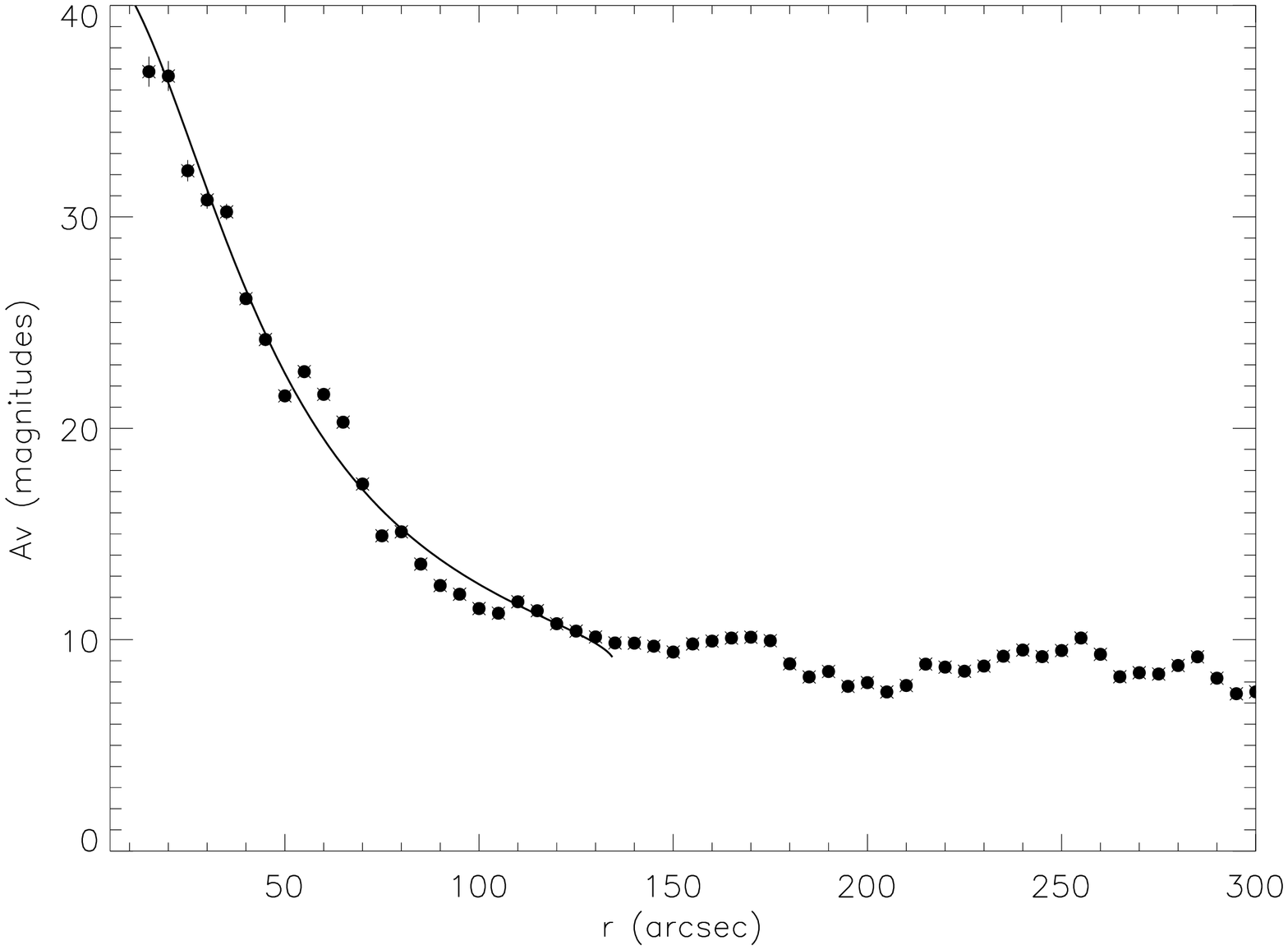}
\includegraphics[angle=90,scale=0.31]{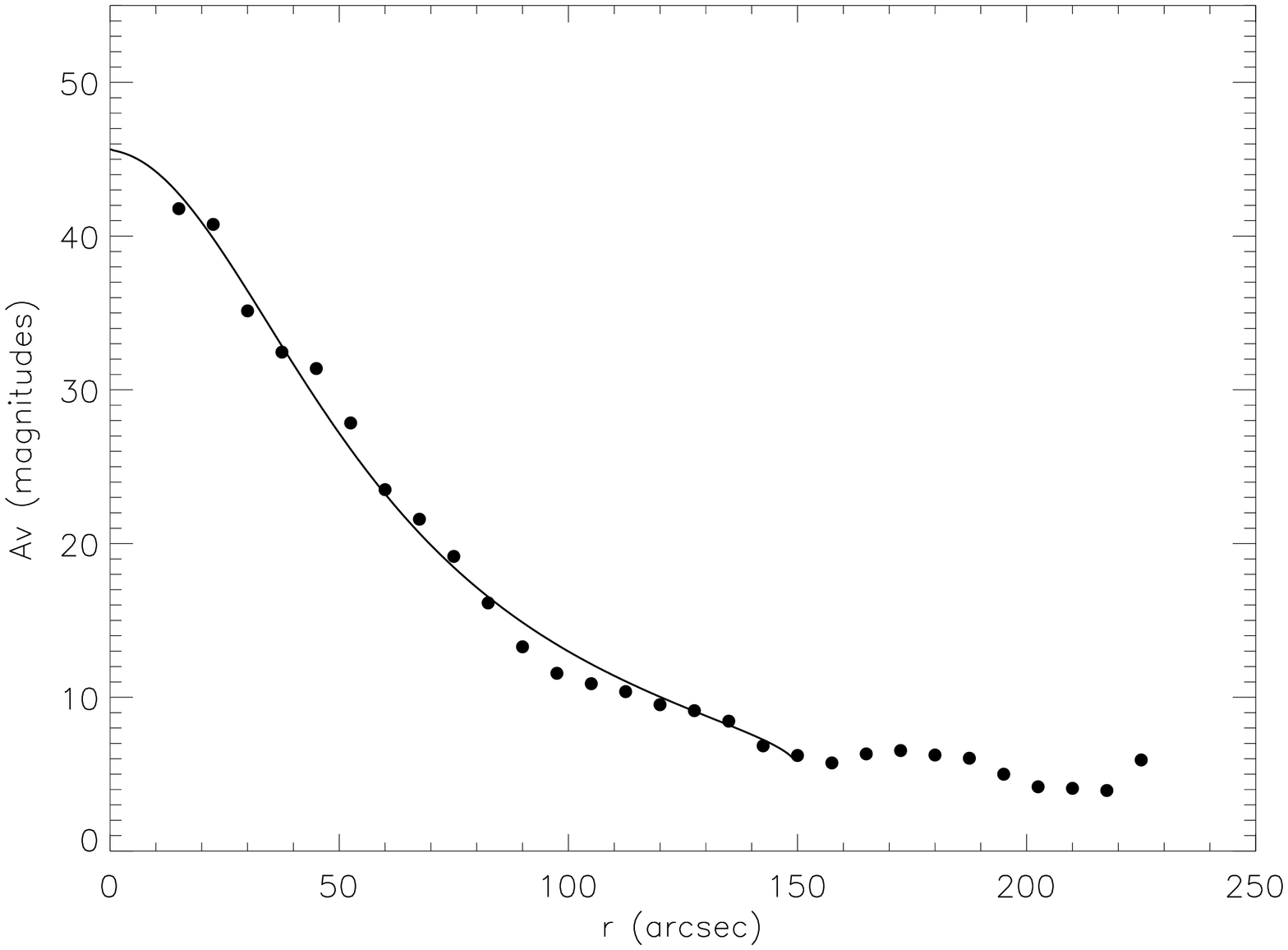}
\caption{Nyquist sampled radial extinction profiles of core C, 10\arcsec\, resolution (left panel) and core D, of lower resolution, 15\arcsec\, (right panel). The overplotted lines corresponds to Bonnor-Ebert fits, discussed in \S\, \ref{subsec:properties}}
\label{fig:CDprofile}
\end{figure}

Figure \ref{fig:coreCDcont} shows a 30\arcsec\, resolution dust extinction map for cores C (West) and D (East). The contours range from 6 to 45 magnitudes and the thicker contour corresponds to 9 magnitudes of extinction. These are two very dense cores that are similar to each other in shape, although core D has higher peak extinction. Core D contains an HH object, \object{HH\,78}, that is a telltale sign of a young stellar object driving an outflow and beginning to clear out surrounding dust. Core C is not known to harbor any young stellar object or HH object. These two cores appear to be connected by a ``bridge'' of gas and dust at a level of 24 magnitudes of extinction. The crosses denote the location of the peak extinctions. Figure \ref{fig:CDprofile} shows the respective profiles for these cores. The peak extinction measured of core C is 38 magnitudes, while it is 42 magnitudes for core D. Both profiles drop until they reach a plateau of 9 magnitudes, giving them the same radii of 135\arcsec.

\begin{figure}
\centering
\includegraphics[angle=90,scale=0.35]{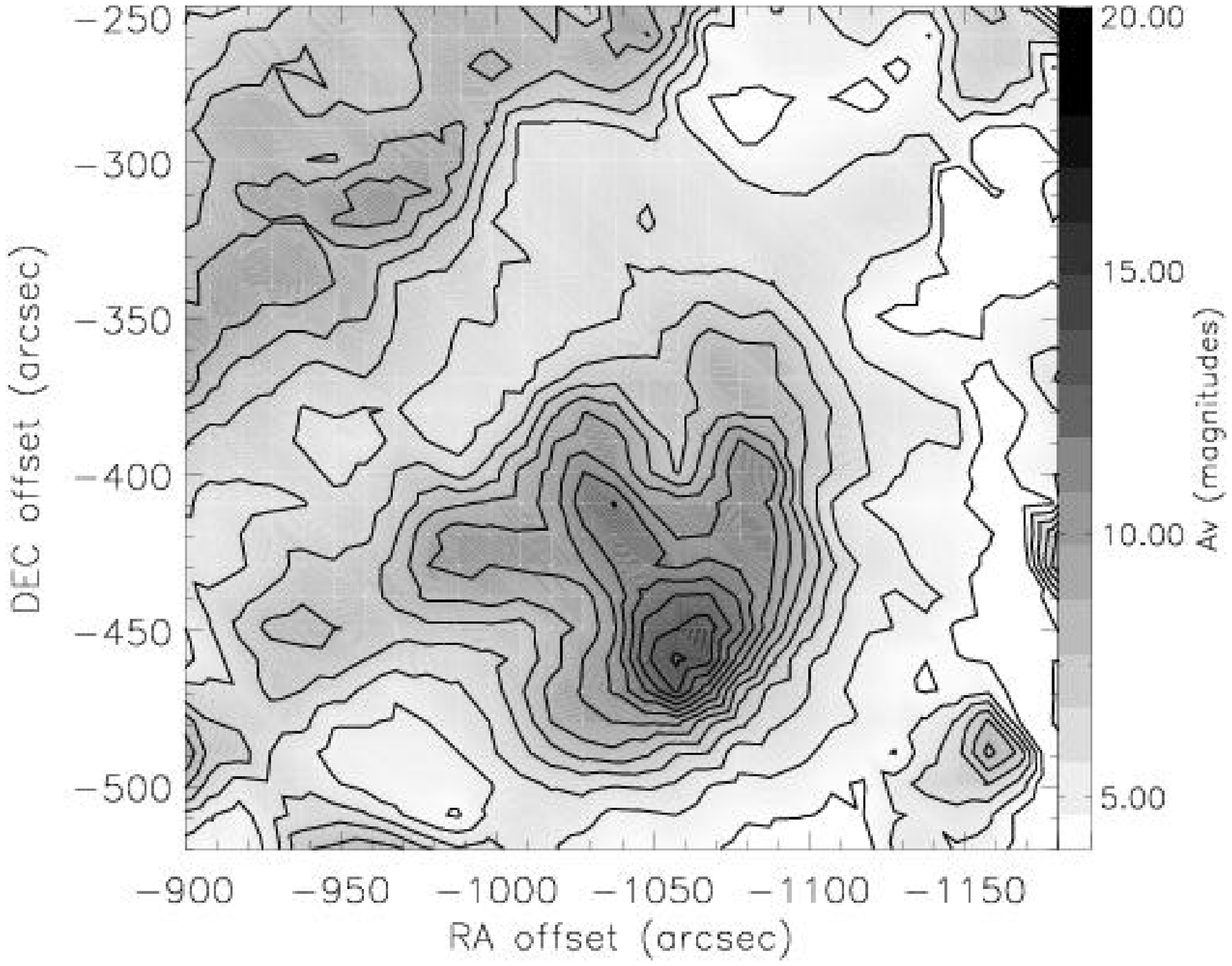}
\includegraphics[angle=90,scale=0.3]{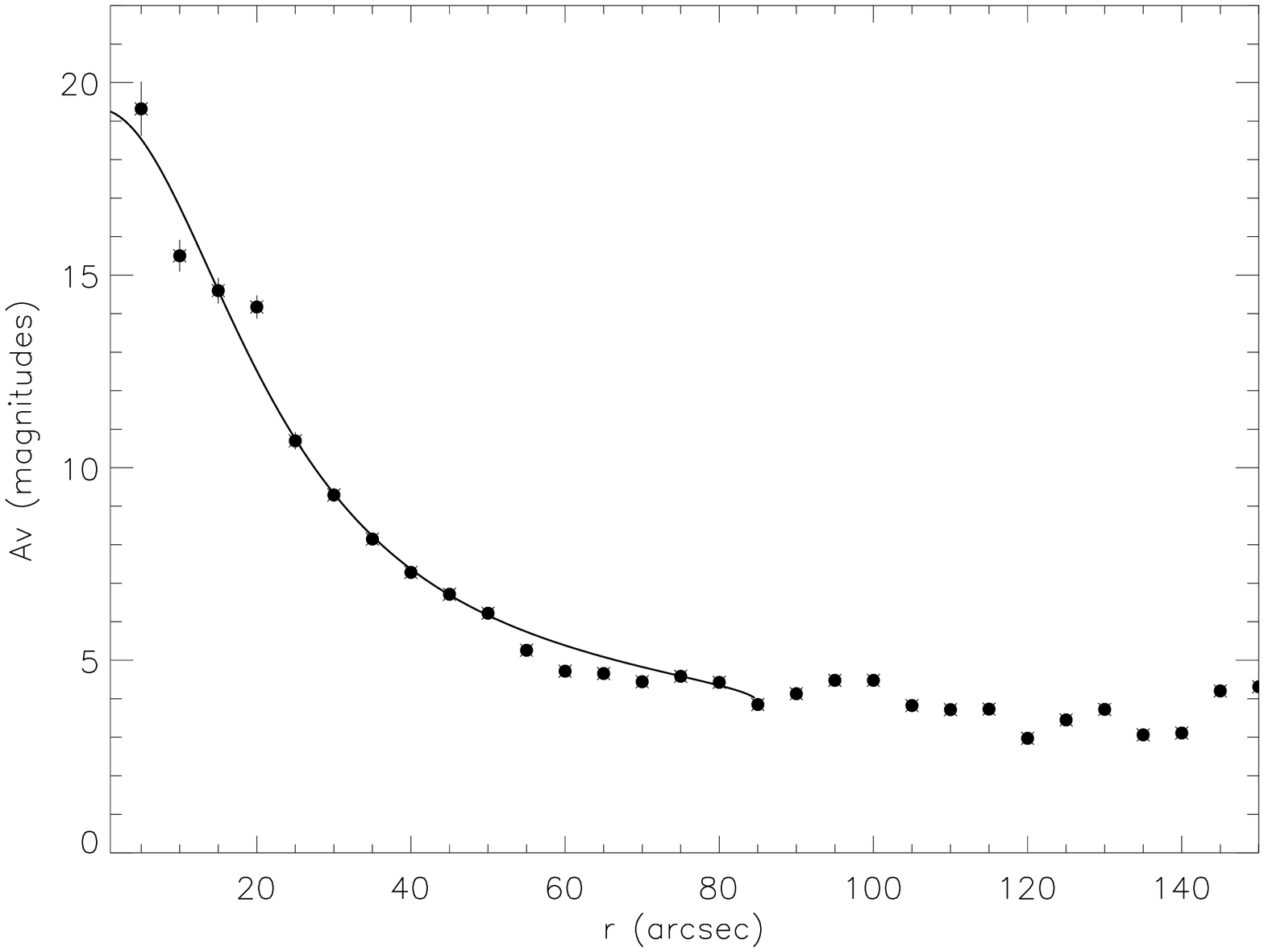}
\caption{\emph{left} Nyquist sampled 20\arcsec\, resolution dust extinction  map of core H. The contours range from 4 to 16 magnitudes in steps of 1 magnitude (color version available in the electronic publication); \emph{right} Nyquist sampled radial extinction profile of core H. The solid line corresponds to a Bonnor-Ebert fit as discussed in \S\ \ref{subsec:properties}}
\label{fig:H}
\end{figure}

The left panel of Figure \ref{fig:H} shows a 20\arcsec\, resolution dust extinction map of core H. The contours range from 4 to 16 magnitudes. This is a small core, compared to the other cores described above. It shows some asymmetry, having a tail that points towards the cluster and is more structured than the other cores. The right panel of this figure shows the radial extinction profile. The peak extinction is measured to be 19.5 magnitudes and the profile gives a radius of 80\arcsec.

\begin{figure}
\centering
\includegraphics[angle=90,scale=0.35]{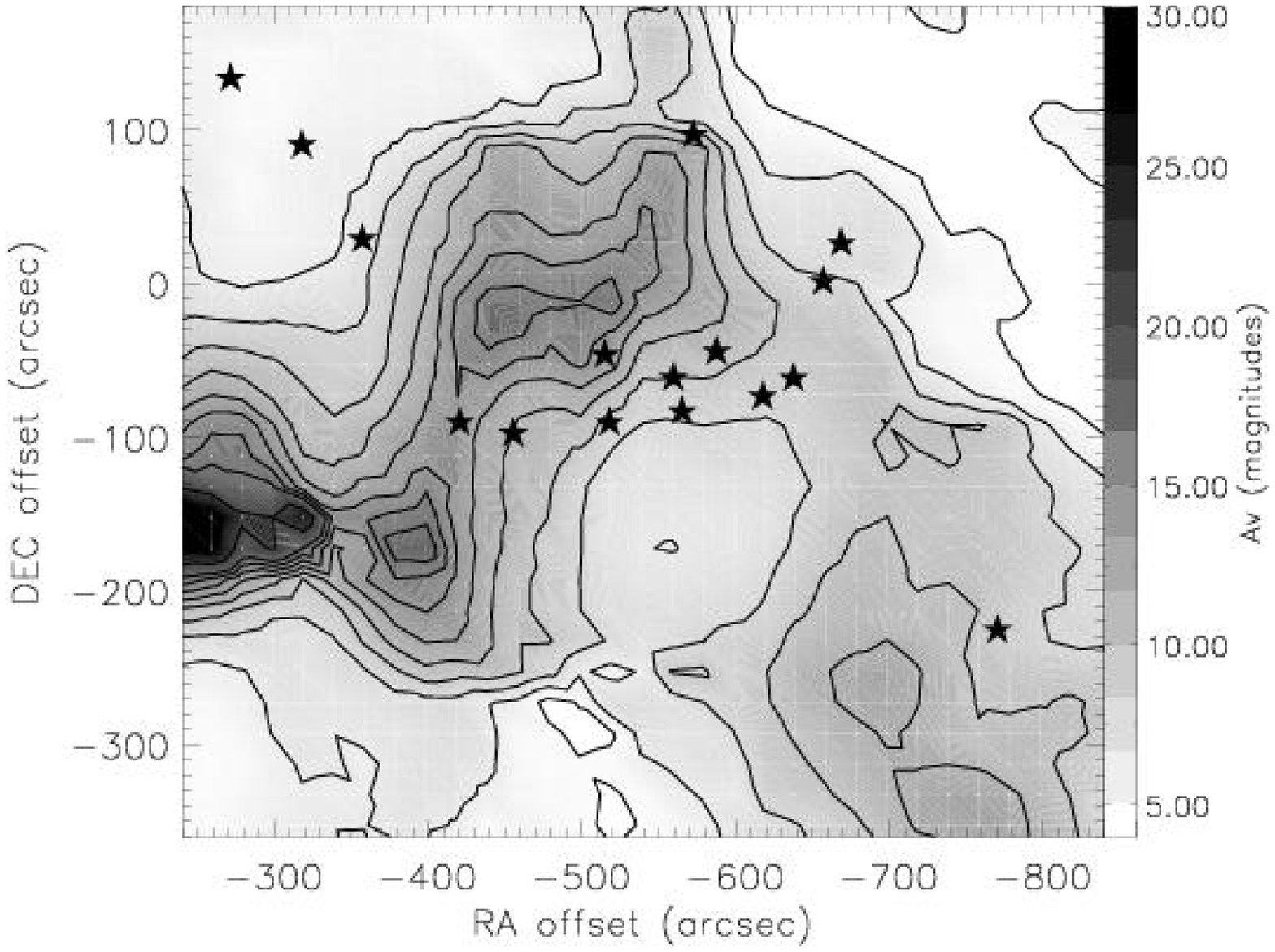}
\includegraphics[angle=90,scale=0.3]{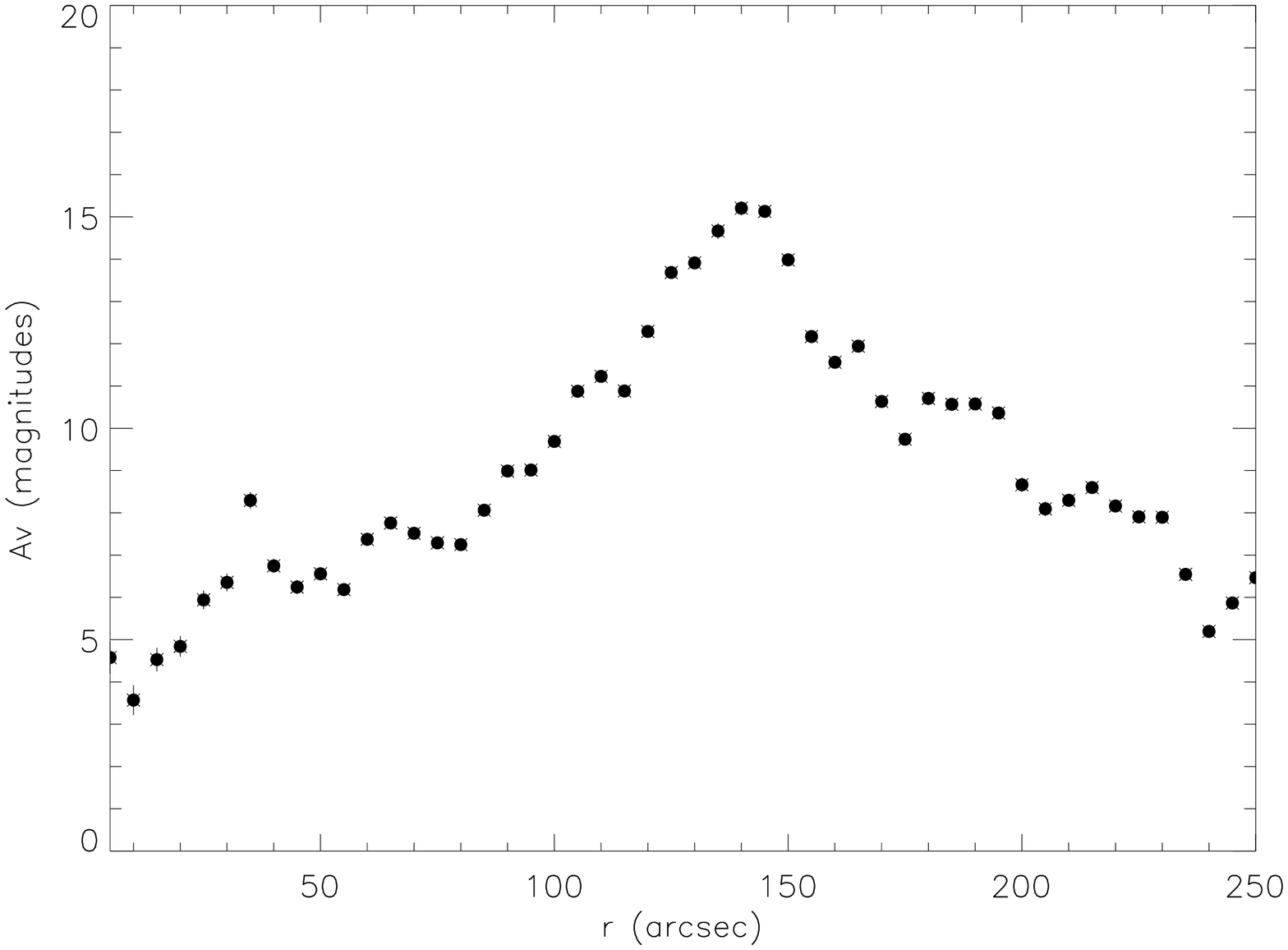}
\caption{\emph{left}: Dust extinction map of the ring structure surrounding the Lupus\,3 cluster. The stars mark the position of H$\alpha$ stars \citep{schwartz}. The contours range from 4 to 20 magnitudes in steps of 2 magnitudes of visual extinction and then to 45 in steps of 5 magnitudes of visual extinction (color version available in the electronic publication); \emph{right}: Radial dust extinction profile of the ring structure bordering the Lupus\,3 cluster, at 10\arcsec\, resolution, and Nyquist sampled.}
\label{fig:ring}
\end{figure}

Figure \ref{fig:ring} shows both a zoomed-in dust extinction map of the ring we identified and its radial profile. The contours range from 4 to 45 magnitudes of equivalent visual extinction. We have overplotted known H$\alpha$ stars to locate the cluster. The peak extinction of the radial profile is 15.5 magnitudes at a radius of 140\arcsec\ which corresponds to 0.11\,pc at the distance assumed for Lupus\,3. The extinction at the center of the ring is approximately 4 magnitudes. It is interesting to note that the sizes of the cores, particularly core C and D, and the ring are very similar.

\begin{figure}
\centering
\includegraphics[angle=90,scale=0.31]{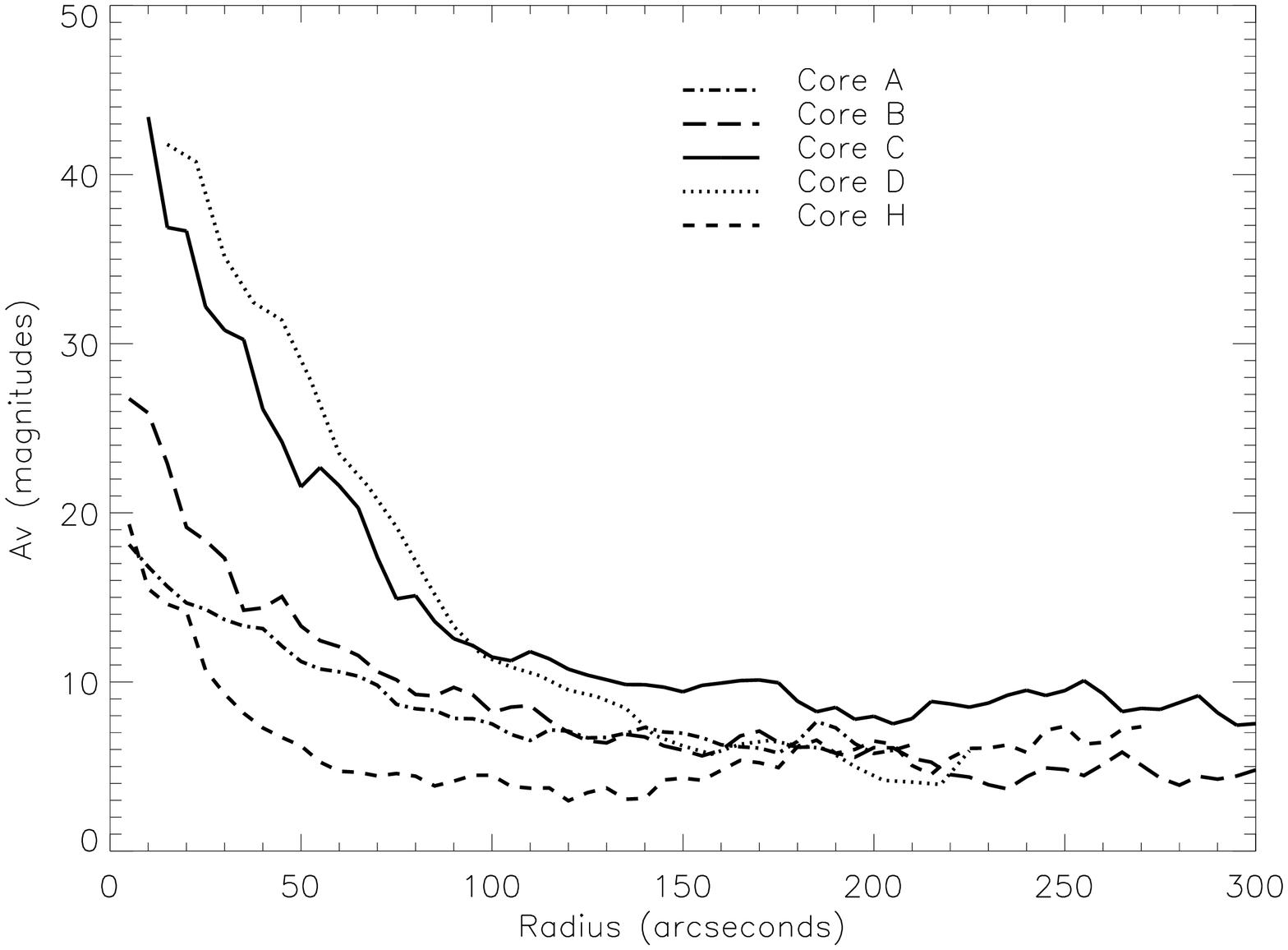}
\includegraphics[angle=90,scale=0.31]{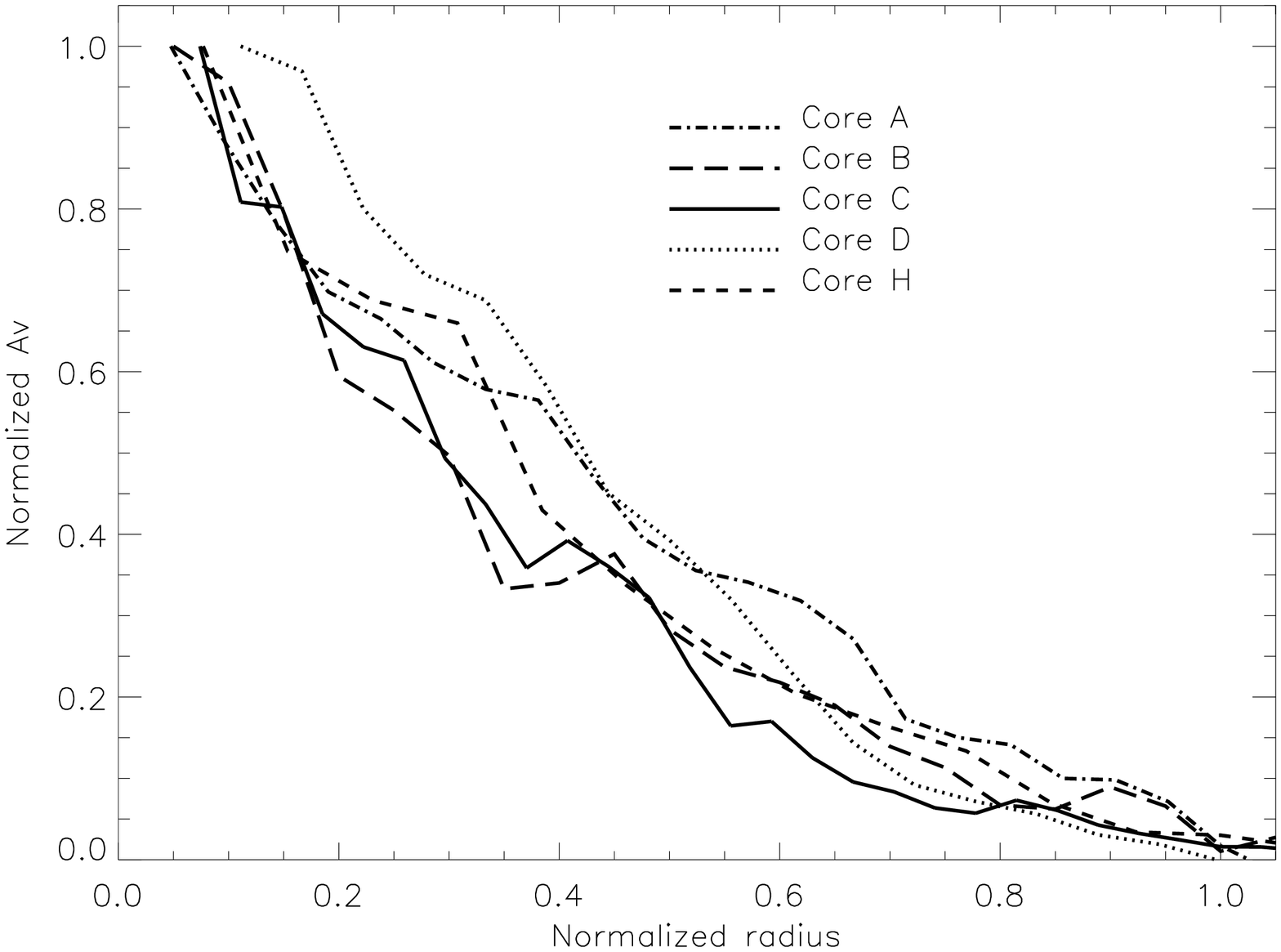}
\caption{\emph{Left}: Comparison of the empirical extinction profiles of all the cores analyzed. \emph{right}: Normalized empirical extinction radial profiles of the cores.}
\label{fig:emp_profiles}
\end{figure}

To compare the density profiles of the cores, we overplot them in the left panel of Figure \ref{fig:emp_profiles}. The profiles clearly have different shapes, peak extinctions and, plateau levels. In Figure \ref{fig:emp_profiles} we compare the normalized radial profiles for the different cores. The normalized profiles appear to behave like those of Bonnor-Ebert spheres therefore we proceed to analyze them with this theoretical model.

\section{Analysis and Discussion}
\label{sec:analysisDiscuss}

\subsection{Physical Properties of the cores}
\label{subsec:properties}

Bonnor-Ebert spheres are self-gravitating, pressure truncated, isothermal spheres in hydrostatic equilibrium, described by the following equation: 
\begin{equation}
\frac{1}{\xi^2}\cdot\frac{d}{d\xi} \left({ \xi^2 \frac{d \psi}{d \xi} }\right)=e^{-\psi}
\label{eq:be}
\end{equation}
where $\xi=(r/a)\sqrt{4\pi\rho_cG}$ is the dimensionless radial parameter, $a=\sqrt{kT/(\mu m_\mathrm{H})}$ is the isothermal sound speed, $k$ is the Boltzmann constant, $\rho_c$ is the volume density at the origin, and $\psi(\xi)=-ln({\rho}/{\rho_c})$. The usual boundary conditions are:
\begin{eqnarray}
\psi(0)=0\\
\frac{d\psi(0)}{d\xi}=0
\end{eqnarray}

The parameter that characterizes the solution to Equation \ref{eq:be} is $\xi_\mathrm{max}$, the value of $\xi$ at the outer boundary. The Bonner-Ebert fit to the profile of each core is shown in the individual profiles (Figures \ref{fig:coreAcont} to \ref{fig:H}). These fits were obtained by choosing the Bonnor-Ebert solution to Equation \ref{eq:be} with the smallest reduced $\chi^2$ \citep[e.g.][]{lada04, l781}. The values of $\xi_\mathrm{max}$ derived are uncertain due to structure in the foreground/background extinction (i.e., plateau level). This situation complicates the precise determination of the radii of the cores. Moreover, our cores are somewhat elongated or flattened and not strictly circular. Nonetheless it is still illustrative to derive basic physical parameters from these fits.
Derived quantities from the Bonnor-Ebert fits to the empirical curves are summarized in Table \ref{tab:comp_cores}. Following the procedure described in \S \ref{subsubsec:mass} for determining the mass of the individual cores, the peak extinction for the BE fitting, A$_V$(max), is also corrected for background/foreground extinction. This peak extinction, together with the radius, $R$, and $\xi_\mathrm{max}$ are used to calculate the Bonnor-Ebert temperature:
\begin{equation}
T_\mathrm{BE}=\frac{R\,A_V(max)}{\kappa(\xi_\mathrm{max})\,(10^8\,\xi_\mathrm{max})^2}\ (K)
\label{eq:temp}
\end{equation}
where 
\begin{equation}
\kappa(\xi_\mathrm{max})=2\,\int^1_0\frac{\rho(r)}{\rho_c}\,d\left(\frac{\xi}{\xi_\mathrm{max}}\right)
\label{eq:kappa}
\end{equation}
is the dimensionless column density \citep[see][]{lada04}. The central density is given by:
\begin{equation}
\rho_c=\frac{1}{4\pi G}\left(\frac{a\,\xi_\mathrm{max}}{R}\right)^2
\label{eq:rhocen}
\end{equation}
where the central number density, $n_c$, is simply $\rho_c/(\mu m_\mathrm{H})$. We use the value of 2.34 for the mean molecular weight $\mu$ after correcting for heavy elements \citep{allen73}.
The external pressure of each core is calculated by \citep{spitzer}:
\begin{equation}
P_\mathrm{ext}=a^2\,\rho_c\,e^{-\psi(\xi_\mathrm{max})}
\end{equation}
This corresponds to the pressure the surrounding cloud material exerts on the individual cores, confining them to their present sizes. The calculated results show that cores A, B and C have the same external gas pressures within a factor of 2. The external pressure derived for core D is considerably larger then the value for the previous cores (see Table \ref{tab:comp_cores}). This suggests that core D may not be in equilibrium and that the Bonnor-Ebert condition may break down in this case. The overall external pressure for the cores ranges from 0.7 to 4.0\,$\times$\,10$^{-12}$\,Pa, or equivalently P$_\mathrm{ext}/k$ ranges from 5.0 to 29.0\,$\times$\,10$^4$\,(cm$^{-3}$\,K). This likely reflects the internal inter-core pressure within the Lupus\,3 cloud. 
We note that if this pressure originates in a turbulent envelope around the cores, the pressure is given by $\rho V^2$, where $\rho$ is the density and $V$ is the turbulent gas velocity. Gas in such an envelope would be expected to have a linewidth of $\Delta V_\mathrm{FWHM} =(2.35/\sqrt{3})V$. For example, densities characteristic of $^{12}$CO emission (n=100\,cm$^{-3}$) would produce an observed linewidth of approximately 2.6\,km\,s$^{-1}$, whereas for densities characteristic of $^{13}$CO emission (n=1000\,cm$^{-3}$), we predict a linewidth of 0.8\,km\,s$^{-1}$. Observations of this area by Thomas Dame (private communication) with the CfA 1.2\,m telescope give a linewidth in $^{12}$CO of 2.8\,km\,s$^{-1}$, and $^{13}$CO observations by \cite{vilas-boas2} give a linewidth of 1.1\,km\,s$^{-1}$, which both agree very well with our predictions.

From the Bonnor-Ebert fitting we are also able to calculate the Bonnor-Ebert mass of the cores, using the following equation \citep{spitzer}:
\begin{equation}
M_\mathrm{BE}=\frac{1}{\sqrt{4\pi \rho_c}}\left(\frac{kT}{\mu m_\mathrm{H}\,G}\right)^{3/2}\xi^2\frac{du}{d\xi}
\end{equation}
These masses are similar to what we derive from integrating the dust column density in each core and are summarized in the last column of Table \ref{tab:comp_cores}.

It may not be strictly valid to use the Bonnor-Ebert sphere model for the analysis of the embedded cores in Lupus\,3, since these appear more elongated than spherical in shape. Nonetheless, the application of this model is extremely useful for it allows us to compare quantitatively the physical parameters of the cores.

\subsection{Core Evolution and Star Formation}
\label{subsec:evolution}

The parameter $\xi_\mathrm{max}$ is related to the center-to-edge density contrast, $\rho_c/\rho_e$ ($\rho_e=\rho(R)$), in a pressure truncated isothermal sphere. For small values of $\xi_\mathrm{max}$, the center-to-edge density contrast is smaller than for large values of $\xi_\mathrm{max}$. The larger value of $\xi_\mathrm{max}$, the steeper a radial profiles appears. \citet{b} and \citet{e} analyzed the stability of Bonnor-Ebert spheres and showed that values of $\xi_\mathrm{max}$ $>\,6.5$ correspond to unstable equilibria. Thus, for objects which are true Bonnor-Ebert equilibrium configurations, $\xi_\mathrm{max}$ may be related to the evolutionary state of a Bonnor-Ebert cloud. High values of $\xi_\mathrm{max}$ and steeply rising density profiles would presumably correspond to unstable regions undergoing star formation while small $\xi_\mathrm{max}$ and relatively shallow density gradients would correspond to stable or perhaps pre-star forming cores.

Indeed, the shapes of extinction derived radial profiles of a few well known globules appear to support this idea. For example, fits of Bonnor-Ebert curves to the radial profiles of the starless cores B\,68 \citep{alves01} and Coalsack G2 \citep{racca,lada04} produce values of $\xi_\mathrm{max}$ between 5 - 6.7, while a fit to the extinction profile of protostellar core B\,355  yields a $\xi_\mathrm{max}$ of 12.5\,$\pm$\,2.6 \citep{b335}. Even though these objects may not be pure Bonnor-Ebert spheres, their density structure appears to be correlated with their status as starless or star-forming cores and is traced by their fitted value of $\xi_\mathrm{max}$.

\begin{figure}
\centering
\includegraphics[angle=90,scale=0.5]{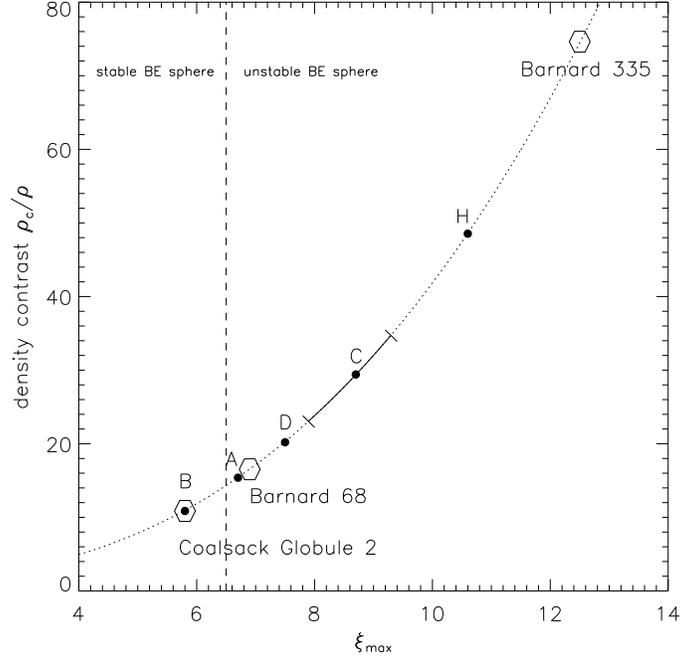}
\caption{Comparison of three different globules of known evolutionary stages (open symbols), Coalsack G2, Barnard 68 and Barnard 335. The dotted line is the Bonnor-Ebert relation between the center-to-edge density contrast and the parameter $\xi_\mathrm{max}$. We also plot the Lupus\,3 cores (solid symbol). The vertical dashed line corresponds to the Bonnor-Ebert critical sphere. For values of $\xi_\mathrm{max}$ less than this the core is likely stable, for greater values that are close to 6.5 the core is assumed to marginally stable, while for values much greater than 6.5 the cores are considered to be unstable if only thermal support is available. The two bars denote the uncertainty in $\xi_\mathrm{max}$ for core C which is representative of the typical error in $\xi_\mathrm{max}$ for the other cores.}
\label{fig:BEplot}
\end{figure}

Figure \ref{fig:BEplot} shows how $\xi_\mathrm{max}$ scales with the density contrast for Bonnor-Ebert spheres\footnote{see also the unpublished result presented by Kandori et al. at the conference \emph{Cores, Disks, Jets \& Outflows in Low \& High Mass Star Forming Environments: Observations, Theory and Simulations}  Banff, Canada, 2004 (\url{http://www.ism.ucalgary.ca/meetings/banff/posters/kandori.pdf})}. On the diagram is  plotted a vertical line that corresponds to the stability threshold for true Bonnor-Ebert equilibrium configurations (i.e., $\xi_\mathrm{max}$=6.5). We also plot on this diagram the locations of the 3 well known globules mentioned previously: Coalsack G2, Barnard\,68 and Barnard\,335. The comparison between these 3 globules clearly illustrates how their evolutionary status is related to the $\xi_\mathrm{max}$ parameter: the more evolved cores have higher density contrasts and higher values of $\xi_\mathrm{max}$. We also plot on this diagram the location of the cores in Lupus\,3 analyzed here. From this plot we are able to draw some general conclusions concerning their relative stability: cores A and B are likely stable and cores D and C are likely unstable, if only thermal support is considered. From this diagram we would also classify core H as being unstable. The stability threshold plotted on this diagram may not necessarily be meaningful for these cores for they may not be bona-fide Bonnor-Ebert spheres dominated primarily by thermal pressure. Nonetheless, regarding  their evolutionary status, cores A and B appear to be the least evolved cores, while cores C and D have the characteristics of being the most evolved cores. 

An independent assessment of the stability of these cores can be obtained by determining their Jeans masses. We derive the Jeans mass of the cores and compare them to the masses calculated from the dust extinction map (see \S \ref{subsubsec:mass}). The ratios between these two values are summarized in Table \ref{tab:masses}, which suggests that cores A, B, and H are Jeans stable, while cores C and D are Jeans unstable. This is consistent with the locations of the cores in Figure \ref{fig:BEplot} and the Bonnor-Ebert analysis, with the exception of core H. We conclude therefore that cores A and B are likely stable, while cores C and D are likely unstable. The status of the stability of core H is not clear given our present data. The conclusions reached regarding the stability of the cores (from either the Bonnor-Ebert or from the Jeans analysis) assume only thermal support against gravitational collapse. C$^{18}$O observations by \citet{vilas-boas2} reveal a relatively narrow linewidth ($0.41\,$km\,s$^{-1}$) for core D. We calculate the ratio of thermal to nonthermal pressure of the core for a range in isothermal sounds speeds corresponding to temperatures between 10 (a temperature typical of starless cores) to 32.8\,K (the Bonnor-Ebert temperature derived for core D). This gives us a range in this ratio of 0.56 to 1.85 which indicates that thermal pressure is a significant source of support. 

\begin{figure}
\centering
\includegraphics[width=10cm]{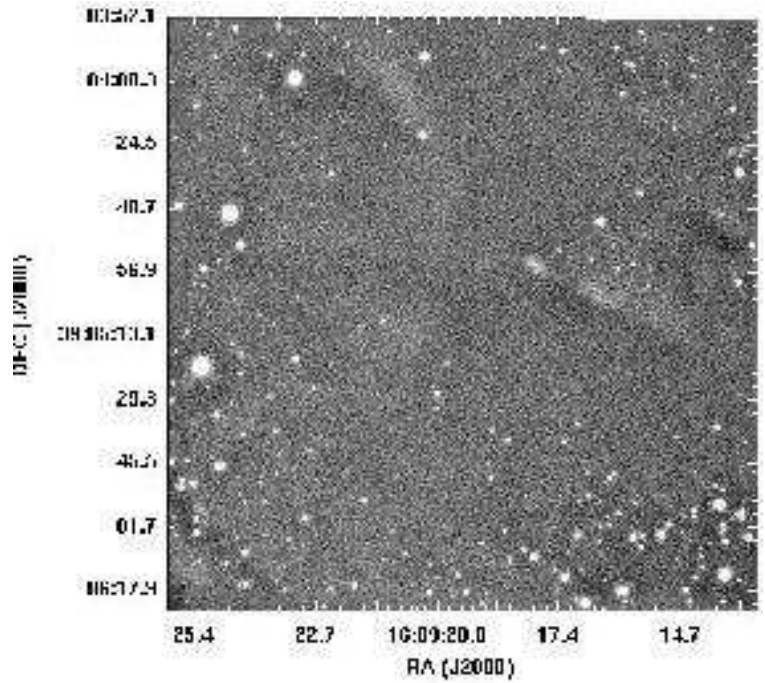}
\caption{Gray scale composite H-band and Ks-band VLT image of the central core C in Lupus\,3. We detect an elongated source at ($\alpha$,$\delta$)$_{J2000}$=(16$^\mathrm{h}$9$^\mathrm{m}$18$^\mathrm{s}$,-39\degr4\arcmin55\arcsec) that could be a jet that created the HH\,78 object \citep{hh78}, located 1\arcmin\, to the west. We do not detect HH78 in our wider FOV image from NTT. The other faint nebulosity, located 42\arcsec\, North-East of the elongated source, may be the outblown cavity around an invisible counter-jet.}
\label{fig:vlt_jet}
\end{figure}

Further examination of the evolutionary status of the Lupus\,3 cores requires a census of star forming activity within the cloud. To pursue this we have searched the literature as well as existing IRAS and \emph{Spitzer} data.
 Cores A and B have no IRAS sources catalogued, and inspection of the \emph{Spitzer} images from the data publicly available from the Cores to Disks \emph{Spitzer} Legacy Program \citep{spitzer03} shows no MIPS sources detected, hence we conclude that cores A and B are likely starless cores. There is an IRAS source located in the filament, IRAS 16067-3902, (see Figure \ref{fig:30map}) and in core C, IRAS 16063-3856, however, these sources have only upper limits to fluxes in all the bands and inspection of the images do not show convincing proof for point source detection so these detections could be spurious. We therefore assume core C, as well as the filament that connects core A to core E, are starless. Further \emph{Spitzer} observations may clarify this issue since core C  may be already collapsing and entering a star-forming state (see Figure \ref{fig:BEplot}). There is evidence for star formation in core D: although the previously known Herbig-Haro object \object{HH\,78} \citep{hh78} is not detected in our images, we clearly see in our VLT $H-$ and $K_s$-band images of core D an elongated source in the southeast which we believe is evidence of a jet associated with HH\,78 and nebulosity in the northeast that would correspond to a cavity around a counter-jet (Figure \ref{fig:vlt_jet}). There is an IRAS source catalogued for this core, \object{IRAS\,16059-3857}, however, detections for this source are marginal given that the cloud is very bright at the longer wavelengths and inspection of the IRAS images does not confirm the presence of a point source. Nonetheless we estimate the bolometric luminosity for this source following \citet{casoli} and get an upper limit of L\,$<$\,0.4\,L$_\sun$.

As mentioned before in \S \ref{subsubsec:profiles}, we note that there is a concentration of H$\alpha$ stars \citep{schwartz} near the dust ring structure, corresponding to the cluster that surrounds the HAe star HR\,5999.
Assuming that the stars in the cluster formed with a 1-D velocity dispersion, $\sigma_V=1.9$\,km\,s$^{-1}$, typical of that characterized by C$^{18}$O observations of this region \citep{vilas-boas2}, then in 0.5\,Myr (the estimated age of HR\,5999) the cluster members would have traveled 0.08\,pc from their place of origin. In Figure \ref{fig:ring} we overplot, on the dust extinction map of the ring, H$\alpha$ stars from \citet{schwartz}. They are concentrated within the ring with only a slight spatial offset between them and the ring center. If we assume that the dust ring corresponds to disrupted material that comprised the original dense core that produced the cluster, we can estimate the star formation efficiency that characterized the formation of the cluster. We estimate a total stellar mass of 11.5 \,M$_\sun$ and thus a star formation efficiency of 30\% for this part of Lupus\,3. This is an upper limit  since star formation, especially that of HR\,5999 and HR\,6000, would have ejected mass from the original core in which they formed by their energetic outflows and it is difficult to estimate how much of this mass was indeed ejected from the parental core. If we were to assume that the efficiency of the parental core was 10\% then that would mean that $\sim$\,100\,M$_\sun$ has been ejected from the core. There is a lot of mass surrounding the remaining denser cores and filament of Lupus\,3, $\sim$\,200\,M$_\sun$ as calculated by \cite{andreazza}, so we cannot exclude the scenario of a considerable amount of mass being expelled into the surrounding region by the star formation in the cluster parental core.
This cluster is likely not gravitationally bound so we expect it to continue to expand and disperse into the surrounding region and contribute to the population of the halo of low-mass stars in the Lupus complex \citep{krautter91,tachihara}.

Overall, there is a spatial gradient of the peak A$_V$ and the peak density of the cores, increasing from east to west. The direction of this gradient coincides with the gradient of the evolutionary phases of the cores (see Figure \ref{fig:BEplot}): the cores on the easternmost side seem less evolved than those of the westernmost side of Lupus\,3. There is also a cluster in what seems to be the remnants of a disrupted massive core on the westernmost side. We interpret this scenario as a sequence of  pre-stellar to  stellar to post-stellar cores within Lupus\,3. If the cores are coeval, as seems likely since all are embedded in the same filamentary cloud, then the core evolution and star formation have proceeded more rapidly for the most massive cores, (i.e., C and D). This leaves open the question of whether the lower mass cores (e.g. A and B) will ultimately evolve to form stars or remain relatively stable until disrupted by other star forming activity in the cloud.

Finally, we note that the cores discussed here are embedded in a filamentary structure and thus have sufficient external pressure to confine them. This raises some interesting questions. Were one of these cores, B for example, to eventually pinch off the filament as the cloud evolves or disrupts, would it maintain its current shape? In such a situation the core could  either expand until it reaches a stable, confined configuration or it may dissolve into the medium if the pressure is too low. Also, if it were to eventually expand until it reached equilibrium, would it become more spherical? Are the elongated shapes often observed for Bok globules a telltale sign that they were once embedded in filamentary clouds such as Lupus\,3?

\section{Summary and conclusions}
\label{sec:summary}
We have performed a deep near-infrared imaging survey of a large portion of the Lupus\,3 dark cloud complex. We have measured the infrared colors and derived extinctions to more than 13,000 stars background to the cloud. We use these data to construct a detailed dust extinction map of the region and examine the structure of the cloud. We briefly summarize the main results of this paper as follows:
\begin{enumerate}
\item{Integrating the column density over the map we find the mass of the cloud to be 84.7$[d(pc)/140]^2$\,$\pm$\,3.9$[d(pc)/140]^2$\,M$_\sun$ within the surveyed region.}
\item{Our observations resolve several embedded cores with different sizes and peak extinctions within the Lupus\,3 cloud. The masses of the cores range from approximately 1 to 8\,M$_\sun$.}
\item{We construct radial column density profiles of the individual cores and fit Bonnor-Ebert profiles to each one. This enables us to estimate certain physical parameters (e.g. central density, external pressure and center-to-edge density contrast) and compare how these vary between cores. We analyze the stability of the cores with both a Jeans and Bonnor-Ebert criteria. Two low extinction and low mass cores (A and B) in the easternmost part of the cloud appear to be relatively stable. Both cores also appear to be starless. Two more massive and higher extinction cores (C and D) in the central region of the cloud appear to be gravitationally unstable and one of these contains an active protostar.}
\item{Overall, there is a gradient from East to West in the peak extinction, which increases towards areas of ongoing and recent star-formation. This corresponds to a gradient in star forming activity of the cloud, from starless/pre-stellar cores in the easternmost part of the cloud, to cores with recent and ongoing star formation in the westernmost part of the cloud, including a likely remnant of a core that harbors a small revealed cluster.} 
\item{From the dust extinction map we also identify a structure in the form of a ring at the westernmost edge of the cloud. A small revealed cluster of the young stellar objects is found projected within the ring, near its northernmost border. We suggest that the ring is the remnant of a dense core within which the small stellar cluster was originally formed. A star forming efficiency for this cluster forming core is estimated to have been $\le$\,30\%, which is consistent with the typical values found by \citet{stareff} for nearby embedded clusters. The further expansion of this emerging cluster will contribute to the halo of low-mass stars surrounding the Lupus clouds.} 

\end{enumerate}
\acknowledgments

We are grateful to an anonymous referee for the constructive comments and suggestions which improved the paper. We thank Gus Muench for helpful advice, Tracy Huard for assisting with the Bonnor-Ebert fitting procedures and Joanna Levine for assistance with the PinkPhot pipeline.
Support for this work was provided by NASA through grants NAG5-9520 and NAG5-13041. P. Teixeira acknowledges support from the Funda\c{c}\~ao para a Ci\^encia e Tecnologia (FCT) Programa Operacional Ci\^encia Tecnologia Inova\c{c}\~ao (POCTI) do Quadro Comunit\'ario de Apoio III, graduate fellowship SFRH/BD/13984/2003, Portugal.
This publication makes use of data products from the Two Micron All Sky Survey, which is a joint project of the University of Massachusetts and the Infrared Processing and Analysis Center/California Institute of Technology, funded by the National Aeronautics and Space Administration and the National Science Foundation. This work is based [in part] on observations made with the Spitzer Space Telescope, which is operated by the Jet Propulsion Laboratory, California Institute of Technology under NASA contract 1407.

\clearpage

\clearpage

\clearpage

\clearpage


\begin{deluxetable}{lcccccccccccccc}
\tabletypesize{\scriptsize}
\centering
\tablecaption{Statistics of the observations.\label{tab:obs}}
\tablewidth{0pt}
\tablecolumns{15}
\tablehead{
\colhead{Field} & \multicolumn{2}{c}{position} & \colhead{} & \multicolumn{3}{c}{Number of stars\tablenotemark{a}}  & \colhead{} & \multicolumn{3}{c}{Completeness limit {(mag.)}} & \colhead{} & \multicolumn{3}{c}{FWHM$^{(\arcsec)}$} \\
\cline{2-3} \cline{5-7} \cline{9-11} \cline{13-15} \\
\colhead{} & \colhead{$\alpha$ (J2000)\tablenotemark{b}} & \colhead{$\delta$ (J2000)\tablenotemark{c}} & \colhead{} & \colhead{$J$} & \colhead{$H$} & \colhead{$K_s$} & \colhead{} & \colhead{$J$} & \colhead{$H$} & \colhead{$K_s$} & \colhead{} & \colhead{$J$} & \colhead{$H$} & \colhead{$K_s$}
}
\startdata
NTT-1   & 16 07 49 & -39 11 15 & & \nodata & 2124 & 1470 & & \nodata & 18.75 & 18.25 & & \nodata & 0.9 & 0.8\\
NTT-2   & 16 08 11 & -39 09 02 & & \nodata & 1590 & 1207 & & \nodata & 18.75 & 18.25 & & \nodata & 1.0 & 0.9\\
NTT-3   & 16 08 34 & -39 05 59 & &    1143 & 1571 &  449 & &   19.75 & 19.25 & 18.75 & &     0.7 & 0.7 & 0.6\\
NTT-4   & 16 08 56 & -39 03 21 & & \nodata & 2420 & 1678 & & \nodata & 18.75 & 18.75 & & \nodata & 0.8 & 0.8\\
NTT-5   & 16 08 55 & -39 08 08 & & \nodata & 1176 & 951  & & \nodata & 18.50 & 18.25 & & \nodata & 1.1 & 0.9\\
NTT-6   & 16 09 22 & -39 03 21 & &  1567   & 1739 & 1363 & &  20.25  & 19.25 & 18.75 & & 0.6     & 0.8 & 0.7\\
NTT-7   & 16 09 22 & -39 08 08 & & \nodata & 1401 & 1070 & & \nodata & 18.75 & 18.25 & & \nodata & 1.0 & 0.9\\
NTT-8   & 16 09 47 & -39 03 21 & & \nodata & 1951 & 1406 & & \nodata & 18.75 & 18.25 & & \nodata & 0.8 & 0.8\\
NTT-9   & 16 09 47 & -39 08 07 & & 1613    & 1704 & 1394 & & 19.25   & 19.25 & 18.75 & & 0.7     & 0.8 & 0.7\\
NTT-10  & 16 10 09 & -39 03 25 & & \nodata & 2151 & 1648 & & \nodata & 19.25 & 18.75 & & \nodata & 0.7 & 0.7\\
NTT-11  & 16 10 06 & -39 11 28 & & \nodata & 2492 & 1774 & & \nodata & 18.75 & 18.75 & & \nodata & 0.6 & 0.6\\
NTT-12  & 16 10 31 & -39 10 48 & & \nodata & 1283 & 925  & & \nodata & 18.75 & 18.25 & & \nodata & 0.8 & 0.9\\
control & 16 08 18 & -38 49 15 & & 2440    & 2339 & 1626 & & 18.25   & 18.25 & 18.25 & & 0.7     & 0.7 & 0.7\\
VLT-1   & 16 09 20 & -39 05 00 & & \nodata &  231 &  472 & & \nodata & 21.25 & 19.75 & & \nodata & 0.6 & 0.4\\
VLT-2   & 16 10 05 & -39 10 43 & & \nodata &  775 &  922 & & \nodata & 21.25 & 19.75 & & \nodata & 0.6 & 0.5\\
\enddata
\tablenotetext{a}{with photometric error $<$\,0.10 magnitudes}
\tablenotetext{b}{given in hours, minutes, and seconds}
\tablenotetext{c}{given in degrees, arcminutes, and arcseconds}
\end{deluxetable}

\clearpage

\begin{deluxetable}{lccccccc}
\tabletypesize{\small}
\centering
\tablecaption{Masses calculated for Lupus\,3 and individual structures therein.\label{tab:masses}}
\tablewidth{0pt}
\tablecolumns{8}
\tablehead{
\colhead{Region} & \colhead{Plateau} & \colhead{Mass\tablenotemark{a}} & \colhead{Mass\tablenotemark{b}} & \colhead{$\bar{n}$\tablenotemark{a}} & \colhead{$\bar{n}$\tablenotemark{b}}  & \colhead{M$_\mathrm{J}$\tablenotemark{b}} & \colhead{M/M$_\mathrm{J}$}\\
\colhead{}  & \colhead{\scriptsize(magnitudes)} & \colhead{\scriptsize(M$_\sun$)}  & \colhead{\scriptsize(10$^4$ cm$^{-3}$)} & \colhead{\scriptsize(10$^4$ cm$^{-3}$)} & \colhead{\scriptsize(10$^4$ cm$^{-3}$)} & \colhead{\scriptsize(M$_\sun$)} & \colhead{} 
}
\startdata
Core A     &  8 &  3.3  & 1.2 & 5.5 & 2.1  & 3.7 & 0.3\\
Core B     &  7 &  4.3  & 1.3 & 6.0 & 1.8  & 4.0 & 0.3\\
Core C     &  9 &  10.4 & 4.9 & 7.8 & 3.6  & 2.8 & 1.8\\
Core D     &  6 &  13.4 & 8.5 & 6.5 & 4.1  & 2.6 & 3.2\\
Core H     &  4 &  2.3  & 0.9 & 6.2 & 2.4  & 3.4 & 0.3\\
Ring       & \nodata & 25 & \nodata & \nodata & \nodata & \nodata & \nodata \\
Lupus\,3 cloud & \nodata  & 84.7  & \nodata  & \nodata & \nodata & \nodata & \nodata \\
\enddata
\scriptsize
\tablenotetext{a}{no plateau subtraction.}
\tablenotetext{b}{with plateau subtraction.}
\tablecomments{The masses calculated assume a distance of 140\,pc to Lupus\,3. If the reader chooses to compare these results for another distance it suffices to multiply these masses by $\left[\frac{d (pc)}{140}\right]^2$.}
\end{deluxetable}

\clearpage

\begin{deluxetable}{cccccccc}
\tabletypesize{\small}
\centering
\tablecaption{Derived properties from Bonnor-Ebert fitting.\label{tab:comp_cores}}
\tablewidth{0pt}
\tablecolumns{8}
\tablehead{
\colhead{Core} & \colhead{A$_V$(max)\tablenotemark{a}} & \colhead{R} & \colhead{$\xi_\mathrm{max}$} & \colhead{T$_\mathrm{BE}$}  & \colhead{n$_c$} & \colhead{P$_\mathrm{ext}$/k} & \colhead{M$_\mathrm{BE}$}\\
  & \colhead{\scriptsize(magnitudes)}  & \colhead{\scriptsize(pc)} &   & \colhead{\scriptsize(K)}  & \colhead{\scriptsize(10$^4$ cm$^{-3})$} & \colhead{\scriptsize(10$^4$\,cm$^{-3}$ K)} & \colhead{\scriptsize(M$_\sun$)}
}
\startdata
A     & 16.3     & 0.07       & 6.7       & 9.5     & 10.6    & 5.8        & 1.3\\
B     & 12.2     & 0.07       & 5.8       & 9.0     & 5.9     & 5.0        & 1.2\\
C     & 33.5     & 0.09       & 8.7       & 18.5    & 20.7    & 10.9       & 3.3\\
D     & 40.2     & 0.10       & 7.5       & 30.6    & 18.8    & 29.0       & 6.4\\
H     & 15.4     & 0.06       & 10.6      & 4.5     & 17.2    & 1.5        & 0.5\\

\enddata
\tablenotetext{a}{Maximum visual extinction given by the Bonnor-Ebert fit}
\end{deluxetable}

\clearpage

\end{document}